\documentclass[a4paper,12pt]{article}

\usepackage{amsmath}
\usepackage{graphicx}

\def\eq#1{\begin{equation}\label{#1}} \def\eeq{\end{equation}}

\newcounter{prim}
\def\prim{\stepcounter{prim}\textit{Example}~\arabic{prim}.~}

\begin{document}
\begin{center}
\uppercase{\textbf{Potentials allowing integration of the perturbed two-body problem 
in regular coordinates}}
\end{center}
\begin {center}
{\sc S.M.~POLESHCHIKOV} \\
{\footnotesize Department of Mathematics, Syktyvkar Forestry Institute, 
Russia,\\ e-mail: polsm@list.ru}


\end {center}
\footnotesize { \noindent
{\bf Abstract.}
The problem of separation of variables in some coordinate systems obtained with the use of $L$-transformations is studied.
Potentials are shown that allow separation of regular variables in a perturbed two-body problem.
The potential contains two arbitrary smooth functions.
An example of a potential is considered allowing explicit solution of the problem in terms of elliptic functions. The cases of bounded and unbounded motion are shown. The results of numerical experiments are given.

\vskip2mm \noindent
{\bf Notation}.
Everywhere below vectors are regarded as column vectors, and are noted
by bold letters. The sign ${}^T$ placed over the vector or matrix symbol
denotes transposition. A quantity evaluated at the initial moment of physical or fictitious time is denoted by zero superscript: $f(0)\equiv f^0$. 
}

\normalsize
\vskip2mm
\begin {center}
{\bf 1. Introduction}
\end {center}

\noindent The integrable cases of motion equations have great practical value. Their significance is determined by the fact that with the help of their solutions one can analyze the motion. In a number of cases integrable problems are used to construct intermediate orbits \cite {aksenov,Fer}.
One of non-trivial examples of integrated systems is the particle motion in a Newtonian field with additional constant acceleration vector. 
This had been investigated earlier by a number of authors \cite{Belec,Kunic,demin} and applied to analysis of space flights with constant jet acceleration. 
In 1970 this problem had been studied using regular coordinates obtained from the KS-matrix \cite {Kirch}.
In contrast to \cite {Kirch}, in \cite {PSM28} integration 
of the same problem was performed in regular coordinates obtained with the use of $L$-transformations. 

In the present work we consider a problem of constructing potentials allowing integration of the equations of motion.
The idea of our approach consists in the following. 
First, a new dynamic system is constructed, having more degrees of freedom than the original one. To do this, an $L$-transformation is applied. The theory of $L $-matrices and their applications is given in  \cite{PSM21, PSM22}.
Using new coordinates, a general potential is selected, allowing separation of variables in the Hamilton - Jacobi equation. 
After this, an inverse transform to original coordinates is performed, using explicit formulas. 
As a basis for selecting general potential with the required integrability property, a well known Stackel theorem is used \cite{Pars}.
This theorem gives necessary and sufficient conditions for separation of variables for orthogonal Hamilton systems, i.e. systems whose Hamiltonian contains only squares of generalized momentums.

Note that separation of variables depends on a choice of a coordinate system. 
We consider here three kinds of coordinate systems: regular, bipolar and spherical. 
The last two systems are introduced in regular coordinates. 
Canonical equations in regular coordinates are constructed using arbitrary $L$-transformations
from the initial canonical motion equations of the perturbed two-body problem. 
The new equations have also orthogonal form and are invariant with respect to  $L$-similarity transforms.
In the nonperturbed case these equations do not have singularity at the attracting center. 
Due to invariance with respect to some perturbing potentials allowing integrability, one can introduce two additional angular parameters.

As a result of this approach the general solution of original system  is represented in parametric form, where fictitious time plays the role of parameter, while the physical time depends on this fictitious time and initial data. 
This sort of integrability is sometimes called 'Sundman integrability' \cite {Kholsh}. 

As an example of integrable case of the perturbed two-body problem the special kind of potential is given. 
In this example the explicit solution of a problem in terms of elliptic functions is expressed, and 
the criterion of bounded  motion  is formulated.

\begin{center}
   \textbf{2. The separation of variables}
\end{center}

Consider the Hamiltonian function of the perturbed two-body problem
\eq{KK0}
H = H({\bf x},{\bf y}) = \frac{1}{2}|{\bf y}|^2 -\frac{\mu}{r} + V,
\quad \mu=\gamma (m+m_0),\>r=|{\bf x}|,
\eeq
where ${\bf x}=(x_1,x_2,x_3)^{T}$ is the position vector of the point of mass $m$ with respect to the point of 
mass $m_0$; ${\bf y}=(y_1,y_2,y_3)^{T}$ is the generalized impulses ($y_i=\dot{x}_i,\>i=1,2,3$); 
$\gamma$ is the gravitational constant; $V = V({\bf x})$ is the perturbed potential. 

For construction of the equations of motion in regular coordinates  we shall need the $L$-transformation $ {\bf z}=L({\bf q}){\bf q} $ generated by the 
$L$-matrix of the fourth order that has the following properties:
\eq{D1}
 L({\bf q})L^{T}({\bf q}) = L^{T}({\bf q})L({\bf q}) = |{\bf q}|^2E  
 \quad \forall \> {\bf q}\in {\bf R}^4,
\eeq
\eq{D2}
 (L({\bf q}){\bf p})_i = (L({\bf p}){\bf q})_i, \quad i=1,\ldots,p,
\eeq
\eq{D3}
 (L({\bf q}){\bf p})_i = - (L({\bf p}){\bf q})_i,\quad i=p+1,\ldots,4
\eeq
$$\forall \> {\bf q}, {\bf p}\in {\bf R}^4.$$
Here $E $ is the unitary matrix. 
The conditions (\ref {D1}) --- (\ref {D3}) simultaneously hold 
only for $p=1 $ or $p=3$. The quantity $p $ is the rank of $L$-transformation. 
The following theorem can be proved \cite{PSM21, PSM22}.

\textbf{THEOREM 1.}  
 \textit{An arbitrary $L$-matrix generating $L$-transformation of rank three,
has the form }   
\eq{F18}
 L({\bf q})=\left(\begin{array}{c}
    {\bf q}^{T}K_1K_4\\ {\bf q}^{T}K_2K_4\\
{\bf q}^{T}K_3K_4 \\ {\bf q}^{T}K_4
 \\ \end{array}\right),
\eeq
\textit{ where orthogonal skew-symmetric matrices $K_1, K_2, K_3, K_4 $ are equal to either}
\eq{F22}
\begin{array}{l}
 K_i=a_{1i} {\cal U}+a_{2i} {\cal V}+a_{3i} {\cal W}, \quad  i= 1,2,3, \\
 K_4 = a_1{\cal X} + a_2{\cal Y} + a_3{\cal Z},
 \end{array}
\eeq
 \textit{or }
\eq{F23}
\begin{array}{l}
 K_i = a_{1i} {\cal X}+a_{2i} {\cal Y}+a_{3i} {\cal Z}, \quad i= 1,2,3, \\
K_4 = a_1{\cal U} + a_2{\cal V} + a_3{\cal W}.
 \end{array}
\eeq                        
\textit{ The triplet of vectors } ${\bf e}_i = (a_{1i}, a_{2i}, a_{3i})^\top$, $i = 1,2,3$,
\textit{forms an orthonormal basis in ${\bf R}^3$, and
${\bf e} = (a_1, a_2, a_3)^\top $ is an arbitrary unitary vector.}

\textit{
Conversely, the arbitrary four skew-symmetric matrices in the form
$(\ref{F22})$ or $(\ref {F23})$ define the $L$-matrix by the formula~$(\ref{F18})$.}

In the formulae (\ref {F22}) and (\ref {F23}) there are the so-called basic skew-symmetric orthogonal matrices 
{\small
{\renewcommand{\arraystretch}{0.9}
$$ \setlength{\arraycolsep}{0.9mm}
 {\cal U}=\left( \begin{array}{rrrr}
   0 & -1 &  0 &  0    \\
   1 &  0 &  0 &  0    \\
   0 &  0 &  0 & -1    \\
   0 &  0 &  1 &  0    \\
 \end{array}\right),\quad
{\cal V}=\left( \begin{array}{rrrr}
   0 &  0 & -1 &  0    \\
   0 &  0 &  0 &  1    \\
   1 &  0 &  0 &  0    \\
   0 & -1 &  0 &  0    \\
 \end{array}\right),\quad
{\cal W}=\left(\begin{array}{rrrr}
      0 &  0 &  0 & -1    \\
      0 &  0 & -1 &  0    \\
      0 &  1 &  0 &  0    \\
      1 &  0 &  0 &  0    \\
 \end{array}\right),
 $$
 }      
{\renewcommand{\arraystretch}{0.9}
$$
\setlength{\arraycolsep}{0.9mm}
{\cal X}=\left( \begin{array}{rrrr}
   0 &  0 & -1 &  0    \\
   0 &  0 &  0 & -1    \\
   1 &  0 &  0 &  0    \\
   0 &  1 &  0 &  0    \\
 \end{array}\right),\quad
{\cal Y}=\left( \begin{array}{rrrr}
   0 & -1 &  0 &  0    \\
   1 &  0 &  0 &  0    \\
   0 &  0 &  0 &  1    \\
   0 &  0 & -1 &  0    \\
 \end{array}\right),\quad
{\cal Z}=\left(\begin{array}{rrrr}
      0 &  0 &  0 & -1    \\
      0 &  0 &  1 &  0    \\
      0 & -1 &  0 &  0    \\
      1 &  0 &  0 &  0    \\
 \end{array}\right).
$$ 
}}      
\normalsize
The matrices $K_i $ are called \textit{generators} of the $L$-matrix.
If $K_1, K_2, K_3, K_4$ are calculated by the formulae (\ref{F22}) 
then $L({\bf q})$ is called {\it the $L$-matrix of first type}, 
otherwise {\it the $L$-matrix of second type}.

We transfer from variables $t $, $x_i $, $y_i $ 
to the new variables  $\tau $, $q_j$, $p_j$ by the formulae  
$$dt=r\,d\tau,$$
\eq{K7}
\left\{\begin{array}{l}
  {\bf x} = \Lambda({\bf q}){\bf q},   \\
 {\bf y} = \frac{\textstyle 1}{\textstyle 2|{\bf q}|^2}\Lambda({\bf q}){\bf p},\quad {\bf q},\>{\bf p} \in {\bf R}^4 \\
\end{array}\right.
\eeq
where  the matrix $\Lambda({\bf q})$ is found from $(\ref {F18})$ by rejection of the fourth line:
$$
\Lambda({\bf q})=\left(\begin{array}{r}
  {\bf q}^{T}K_1K_4\\ {\bf q}^{T}K_2K_4\\ {\bf q}^{T}K_3K_4 \\  \end{array}\right).
$$

Consider the equations of motion in new variables $q_i$, $p_i$
\eq{K8}
\frac{dq_j}{d\tau} = \frac{\partial {\cal K}}{\partial p_j},\quad
\frac{dp_j}{d\tau} = -\frac{\partial {\cal K}}{\partial q_j},
\quad j=0,1,2,3,4
\eeq
\noindent with the Hamiltonian
\eq{K88}
{\cal K} = \frac{1}{8}|{\bf p}|^2 + p_0|{\bf q}|^2 + |{\bf q}|^2V_c({\bf q}),\quad
V_c({\bf q}) = V\bigl({\bf x}({\bf q})\bigr). \eeq
In this system the first equation with $j=0 $ corresponds to transformation of time:
$dq_0 = |{\bf q}|^2d\tau $. 
The variable $p_0$ is conjugate to $q_0$ and has a constant value.

If 
\eq{K9} {\bf x}(0)={\bf x}^0,\quad {\bf y}(0)={\bf y}^0      \eeq 
are initial conditions for the variables of the system with the Hamiltonian (\ref {KK0}),
then, as it is proved in \cite {KS,PSM13}, with the initial values defined by formulae
\eq{K10}
\left\{\begin{array}{ll}
q_0(0) = 0, & {\bf x}^0 = \Lambda({\bf q}^0){\bf q}^0,   \\
p_0(0) = -H({\bf x}^0,{\bf y}^0), & {\bf p}^0 =  2\Lambda^{T}({\bf q}^0){\bf y}^0,  \\
\end{array}\right.
\eeq
the solution of (\ref {K8}) becomes, under the transformation (\ref {K7}), 
a solution of  the system with the Hamiltonian (\ref {KK0}) satisfying the initial conditions (\ref {K9}).
The function ${\bf q}^{T}K_4{\bf p}$ preserves a constant value 
along solutions of (\ref {K8}), and with the initial conditions from (\ref {K10}), this value is zero \cite {PSM13}. 
Hence, the equality ${\bf q}^{T}K_4{\bf p}=0$ is the first integral of this system. The variable $q_0 $ coincides with physical time $t$.

Note that the systems with Hamiltonian (\ref{KK0}) and (\ref {K88}) have different orders. 
The choice of initial values by the formulae (\ref {K10}) means that 
there is a special construction of the system (\ref {K8})
for each trajectory of the system with Hamiltonian (\ref{KK0}). 

Let's pick up the form of potential $V$, admitting division of variables. 
For this purpose we shall take advantage of the theorem proved by Stackel \cite {Pars}.

\textbf{THEOREM 2.}  
\textit{The system with Hamiltonian} 
$$
H=\sum_{i=1}^{n}c_i(q_1,\ldots,q_n)\Bigl(\frac12p_i^2+V_i(q_i)\Bigr),
$$ 
\textit{admits separation of variables in the Hamilton - Jacobi equation if and 
only if there is a nonspecial matrix $\Phi$ of order $n$ wose elements $ \varphi_{si}$ depend only on $q_i$, such that}
\eq{h14}
\Phi \textbf{c}=(1,0,\ldots,0)^{T},
\eeq
\textit{where} $\textbf{c}=(c_1,c_2,\ldots,c_n)^{T}$.

In this case the integrals of motion will be
\eq{h14b}
\begin{array}{c}
t-\beta_1=\sum\limits_{i=1}^{n}
\int\limits_{q_{i}^0}^{q_i}\frac{\textstyle \varphi_{1i}(q_i)dq_i}{\textstyle \sqrt{f_i(q_i)}},\quad 
-\beta_s=\sum\limits_{i=1}^{n}
\int\limits_{q_{i}^0}^{q_i}\frac{\textstyle \varphi_{si}(q_i)dq_i}{\textstyle \sqrt{f_i(q_i)}},\quad s=2,\ldots,n,\\
p_i=\sqrt{f_i(q_i)},\quad i=1,\ldots,n,
 \end{array}\eeq
where $f_i(q_i)=2(\alpha_1\varphi_{1i}(q_i)+\ldots+\alpha_n\varphi_{ni}(q_i)-V_i(q_i))$; $\alpha_i$, $\beta_i$ $(i=1,\ldots,n)$ is constant. As $q_{i}^0 $ a simple root of the function $f_i (q_i)$ is taken. 

Consider again the separation of variables in regular coordinates $q_i$. 
The Hamiltonian looks like (\ref {K88}).
In this case we have
$$
c_1=c_2=c_3=c_4=\frac14,\quad |{\bf q}|^2(p_0 + V_c({\bf q}))=\frac14\sum_{s=1}^{4}V_s(q_s).
$$
The solution of system (\ref {h14}) will be, for example, the matrix
$$\Phi=\left(\begin{array}{rrrr}
      4 &  0 &  0 &  0    \\
     -1 &  1 &  0 &  0    \\
      0 & -1 &  1 &  0    \\
      0 &  0 & -1 &  1    \\
 \end{array}\right).
$$
The potential is defined up to a constant. As $p_0 $ is a constant, we obtain
\eq{h15a}
  V_c({\bf q})=\frac{1}{4|{\bf q}|^2}(V_1(q_1)+V_2(q_2)+V_3(q_3)+V_4(q_4)).
\eeq

Let's find expression for the potential $V_c({\bf q}) $ in original coordinates $\textbf{x}$. 
Let's notice that variables $x_i $ and $r $ are quadratic forms of the variables $q_1$, $q_2$, $q_3$, $q_4$. 
Using the $L$-similarity transformation it is possible to choose an $L$-matrix such that a linear combination 
$B_1x_1+B_2x_2+B_3x_3 $ it will be equal to the sum of squares of $q_i$ with some coefficients. 
Note that for any $L$-matrix we have $r = |{\bf q}|^2$. 
As $V_c({\bf q})$ is to be of the form (\ref {h15a}), the required potential in $x $-coordinates will be the function of the form
 \eq{h16}   V({\bf x})= \frac{1}{r}(Ar+B_1x_1+B_2x_2+B_3x_3).\eeq

Let's specify a choice of $L$-matrix with the required property. Introduce the notation 
$$B=\sqrt{B_1^2+B_2^2+B_3^2},\quad b_i=\frac{B_i}{ B},\quad i=1,2,3. $$

Suppose that the $L$-matrix is of the first type. 
That is, $K_1 $, $K_2 $, $K_3 $ are calculated by the formula (\ref {F22});
for simplicity we assume that $K_4=-{\cal Y}$. Then
$$ Ar+ B(b_1x_1+b_2x_2+b_3x_3) =Ar-  B{\bf q}^{T}
\Bigl[(b_1a_{11}+b_2a_{12}+b_3a_{13}){\cal U}+$$
$$+(b_1a_{21}+b_2a_{22}+b_3a_{23}){\cal V}+
(b_1a_{31}+b_2a_{32}+b_3a_{33}){\cal W}\Bigr]{\cal Y}{\bf q}.$$
Choose the parameters $a_{ij}$ of $L$-matrix in such a way that the following equalities hold:
\eq{K11}
\left\{\begin{array}{l}
b_1a_{11}+b_2a_{12}+b_3a_{13} = 1,  \\
b_1a_{21}+b_2a_{22}+b_3a_{23} = 0,  \\
b_1a_{31}+b_2a_{32}+b_3a_{33} = 0.  \\
\end{array}\right.
\eeq
Geometrically, the solution to this system means that the vector 
${\bf i}_1=(a_{11},a_{12},a_{13})^{T}$ coincides with ${\bf b}=(b_{1},b_{2},b_{3})^{T}$, and the vectors 
${\bf i}_2=(a_{21},a_{22},a_{23})^{T}$, ${\bf i}_3=(a_{31},a_{32},a_{33})^{T}$ 
are orthogonal to ${\bf b}$. 
Moreover, it follows from the structure of the $L$-matrix 
that vectors ${\bf i}_1$, ${\bf i}_2$,  and ${\bf i}_3$ form a frame. 
It is evident that the system (\ref{K11}) has infinite number of solutions. 
We write its general solution. For the first vector we have
$${\bf i}_1=(b_1, b_2, b_3)^{T}.$$
For ${\bf i}_2$ and ${\bf i}_3$ we assume, in the case $b_1^2+b_2^2 \neq 0$, that
$${\bf i}_2=\frac{1}{\sqrt{b_1^2+b_2^2}}\Bigl(
b_2\cos{\alpha}+b_1b_3\sin{\alpha},\>-b_1\cos{\alpha}+b_2b_3\sin{\alpha},\>
-(b_1^2+b_2^2)\sin{\alpha}\Bigr)^{T},$$
$${\bf i}_3=\frac{1}{\sqrt{b_1^2+b_2^2}}\Bigl(
-b_2\sin{\alpha}+b_1b_3\cos{\alpha},\>b_1\sin{\alpha}+b_2b_3\cos{\alpha},\>
-(b_1^2+b_2^2)\cos{\alpha}\Bigr)^{T}.$$
If $b_1^2+b_2^2 = 0$, then ${\bf b}=(0,0,b_3)^{T}$, $b_3=\pm 1$. Therefore,     
we can take the following vectors as the general solution of the system (\ref{K11}):
$${\bf i}_1=(0, 0, b_3)^{T},\quad
{\bf i}_2=\Bigl(\cos{\alpha},\>\sin{\alpha},\>0\Bigr)^{T},\quad
{\bf i}_3=b_3\Bigl(-\sin{\alpha},\>\cos{\alpha},\>0\Bigr)^{T}.$$
The quantity $\alpha\in [0,\>2\pi]$ plays the role of an arbitrary	
parameter of the general solution.

After choosing the parameters $a_{ij}$, the matrix $\Lambda({\bf q})$ is
determined uniquely. The solution of (\ref {K11}) gives 
$$ V_c({\bf q})=\frac{1}{|{\bf q}|^2}(A|{\bf q}|^2-  B{\bf q}^{T}{\cal U}{\cal Y}{\bf q})=$$
$$=\frac{1}{|{\bf q}|^2}\bigl((A+ B)q_1^2+(A+ B)q_2^2+(A- B)q_3^2+(A- B)q_4^2\bigr).$$
Hamiltonian in $q$-coordinates corresponding to this potential  becomes
$$
{\cal K} = \sum_{i=1}^{4}\frac14\Bigl(\frac12p_i^2+4p_0q_i^2+4D_iq_i^2)\Bigr),
$$
where $D_1=D_2=A+ B$, $D_3=D_4=A- B$. 
The canonical system of the equations falls into four subsystems
\eq{h16a}
\frac{dq_i}{d\tau}=\frac14p_i,\quad \frac{dp_i}{d\tau}=-2(p_0+D_i)q_i,\quad i=1,2,3,4.
\eeq
These systems are equivalent to four harmonious oscillators.
Integrals of motion are obtained either from (\ref {h14b}), or straightforward from solving (\ref {h16a}).
Thus, separation of variables for potential (\ref {h16}) is carried out.

For regular $q$-coordinates, we introduce a new coordinate system.  
To preserve the canonical form of equations of motion, we use the canonical transformation with generating function
$$\Psi = p_1\sqrt{Q_1}\cos{Q_2} + p_2\sqrt{Q_1}\sin{Q_2}+p_3\sqrt{Q_3}\cos{Q_4} + p_4\sqrt{Q_3}\sin{Q_4}.$$
We obtain
\eq{K14}
\begin{array}{l}
q_1=\frac{\textstyle\partial \Psi}{\textstyle\partial p_1}=\sqrt{Q_1}\cos{Q_2},\quad
q_2=\frac{\textstyle\partial \Psi}{\textstyle\partial p_2}=\sqrt{Q_1}\sin{Q_2},\\
q_3=\frac{\textstyle\partial \Psi}{\textstyle\partial p_3}=\sqrt{Q_3}\cos{Q_4},\quad 
q_4=\frac{\textstyle\partial \Psi}{\textstyle\partial p_4}=\sqrt{Q_3}\sin{Q_4},\\
P_1=\frac{\textstyle\partial \Psi}{\textstyle\partial Q_1}=
\frac{\textstyle 1}{\textstyle 2\sqrt{Q_1}}(p_1\cos{Q_2}+p_2\sin{Q_2}),\\ 
P_2=\frac{\textstyle\partial \Psi}{\textstyle\partial Q_2}=
\sqrt{Q_1}(-p_1\sin{Q_2}+p_2\cos{Q_2}),\\
P_3=\frac{\textstyle\partial \Psi}{\textstyle\partial Q_3}=
\frac{\textstyle 1}{\textstyle 2\sqrt{Q_3}}(p_3\cos{Q_4}+p_4\sin{Q_4}),\\
P_4=\frac{\textstyle\partial \Psi}{\textstyle\partial Q_4}=
\sqrt{Q_3}(-p_3\sin{Q_4}+p_4\cos{Q_4}).\\
\end{array}
\eeq
The coordinates $Q_1 $, $Q_2 $, $Q_3 $, $Q_4$, obtained from (\ref {K14}), will be called bipolar.
From the last four equations we find $p_1$, $p_2$, $p_3$, $p_4$:
\eq{K15}
\begin{array}{l}
p_1=2P_1\sqrt{Q_1}\cos{Q_2}-\frac{\textstyle P_2}{\textstyle \sqrt{Q_1}}\sin{Q_2},\quad
p_2=2P_1\sqrt{Q_1}\sin{Q_2}+\frac{\textstyle P_2}{\textstyle \sqrt{Q_1}}\cos{Q_2}, \\
p_3=2P_3\sqrt{Q_3}\cos{Q_4}-\frac{\textstyle P_4}{\textstyle \sqrt{Q_3}}\sin{Q_4}, \quad
p_4=2P_3\sqrt{Q_3}\sin{Q_4}+\frac{\textstyle P_4}{\textstyle \sqrt{Q_3}}\cos{Q_4}. \\
\end{array}
\eeq
In the new variables the Hamiltonian ${\cal K}$ becomes
$$\overline{\cal K}=\frac{1}{8}\Bigl(4Q_1P_1^2+\frac{P_2^2}{Q_1}+4Q_3P_3^2+\frac{P_4^2}{Q_3}\Bigr)+
p_0(Q_1+Q_3)+(Q_1+Q_3)\overline{V},$$
where function $\overline {V}$ is expressed in terms of $Q_i$.

Similar to the above, consider separation of variables in bipolar coordinates. 
In the notations of theorem 2 we now have 
$$
c_1=Q_1,\quad c_2=\frac{1}{4Q_1},\quad c_3=Q_3,\quad c_4=\frac{1}{4Q_3}.
$$
As a solution to (\ref {h14}) one can take the matrix
\eq{h151}\Phi=\left(\begin{array}{rrrr}
      \frac{\textstyle 1}{\textstyle Q_1}    &  0 &  0 &  0    \\
     -\frac{\textstyle 1}{\textstyle 4Q_1^2} &  1 &  0 &  0    \\
     -\frac{\textstyle 1}{\textstyle Q_1}    &  0 &  \frac{\textstyle 1}{\textstyle Q_3} &  0    \\
      0 &  0 & -\frac{\textstyle 1}{\textstyle 4Q_3^2} &  1    \\
 \end{array}\right).
\eeq
For the potential $ \overline {V} $ admitting separation of variables, we find
$$
\overline{V}=\frac{1}{Q_1+Q_3}\Bigl(Q_1\overline{V}_1(Q_1)+\frac{1}{4Q_1}\overline{V}_2(Q_2)+Q_3\overline{V}_3(Q_3)+\frac{1}{4Q_3}\overline{V}_4(Q_4)\Bigr).
$$
In $q$-coordinates we obtain the form
$$
V_c=\frac{1}{|{\bf q}|^2}\Bigl((q_1^2+q_2^2)\overline{V}_1(q_1^2+q_2^2)+
\frac{\overline{V}_2(\arctan\frac{q_2}{q_1})}{4(q_1^2+q_2^2)}+
(q_3^2+q_4^2)\overline{V}_3(q_3^2+q_4^2)+\frac{\overline{V}_4(\arctan\frac{q_4}{q_3})}{4(q_3^2+q_4^2)}\Bigr).
$$
Passing to $x$-coordinates, we use the concrete $L$-transformation 
\eq{h19}
 \left\{ \begin{array}{rl}
 x_1=&\hbox{ } 2q_1q_4+2q_2q_3,\\
 x_2=&-2q_1q_3+2q_2q_4,\\
 x_3=& q_1^2+q_2^2-q_3^2-q_4^2,\\
\end{array} \right.        \eeq
which follows from (\ref {F18}), (\ref {F22}) with $K_1 ={\cal V}$, $K_2={\cal W}$, $K_3={\cal U}$, $K_4=-{\cal Y}$.
Taking into account that for any $L$-matrix the equality $r = q_1^2+q_2^2+q_3^2+q_4^2 $ holds,
we obtain
$$
q_3^2+q_4^2 = \frac{1}{2}(r-x_3), \quad q_1^2+q_2^2 = \frac{1}{2}(r+x_3).      $$
The general solution of the first equation is
$$
q_3 =\sqrt{\frac{r-x_3}{2}}\cos{\psi},\quad
q_4 =\sqrt{\frac{r-x_3}{2}}\sin{\psi},\quad \psi\in [0,\>2\pi].
$$
Then
$$ q_1=\frac{x_1\sin{\psi}-x_2\cos{\psi}}{\sqrt{2}\sqrt{r-x_3}},\quad
 q_2=\frac{x_1\cos{\psi}+ x_2\sin{\psi}}{\sqrt{2}\sqrt{r-x_3}}.  $$
In a similar way we may introduce a parameter, using the second equation,
$$
q_1 =\sqrt{\frac{r+x_3}{2}}\cos{\psi_1},\quad
q_2 =\sqrt{\frac{r+x_3}{2}}\sin{\psi_1},\quad \psi_1\in [0,\>2\pi].
$$
As is well known \cite {KS}, under $L $-transformation for a point in $ \textbf{R}^3 $ at a distance $r $ from origin, there corresponds a point of some circle of radius $ \sqrt{r} $ in $ \textbf{R}^4 $. The variables $q_i $ contain an arbitrary parameter $\psi$ (or $\psi_1$), giving parametrization of the given circle. In the original coordinates $x_i $ this parameter disappears.
Note that 
$$\frac{ q_2}{ q_1}=\tan\psi_1,\quad \frac{ q_4}{ q_3}=\tan\psi. $$
We therefore assume functions $V_2$, $V_4$ to be constant.
Then we arrive at a potential of the form
\eq{KK1}
V({\bf x}) =\frac{1}{r}\Bigl[G_1((r+ x_3)/2)+G_2((r-x_3)/2)\Bigr],
\eeq
where $G_1 $, $G_2 $ are arbitrary smooth functions.
The Hamiltonian in bipolar coordinates for this potential takes the form
$$\overline{\cal K}=
Q_1\Bigl(\frac{P_1^2}{2}+p_0+\frac{G_1(Q_1)}{Q_1}\Bigr)+
\frac{1}{4Q_1}\frac{P_2^2}{2}+Q_3\Bigl(\frac{P_3^2}{2}+p_0+\frac{G_2(Q_3)}{Q_3}\Bigr)+
\frac{1}{4Q_3}\frac{P_4^2}{2}.
$$
In view of the solution (\ref {h151}) for $f_i$ from the theorem 2 we have
$$f_1(Q_1)=2\Bigl(\frac{\alpha_1}{Q_1}-\frac{\alpha_2}{4Q_1^2}-\frac{\alpha_3}{Q_1}-p_0-\frac{G_1(Q_1)}{Q_1}\Bigr),
\quad f_2(Q_2)=2\alpha_2,$$
$$f_3(Q_3)=2\Bigl(\frac{\alpha_3}{Q_3}-\frac{\alpha_4}{4Q_3^2}-p_0-\frac{G_2(Q_3)}{Q_3}\Bigr),\quad
f_4(Q_4)=2\alpha_4.$$
Then integrals of motion are obtained by formulas (\ref {h14b}).

Let's consider one more case of separation of variables. Introduce in $q$-coordinates the spherical coordinates
\eq{KK14}
\begin{array}{l}
q_1=\sqrt{Q_1}\cos{Q_2}\cos{Q_4},\quad
q_2=\sqrt{Q_1}\sin{Q_2}\cos{Q_4},\\
q_3=\sqrt{Q_1}\cos{Q_3}\sin{Q_4},\quad 
q_4=\sqrt{Q_1}\sin{Q_3}\sin{Q_4}.\\
\end{array}
\eeq
We supplement the transformation (\ref {KK14}) to obtain a canonical transformation of impulses 
\eq{KK14a}
\begin{array}{l}
p_1=2\sqrt{Q_1}\cos{Q_2}\cos{Q_4}P_1 -\frac{\textstyle  \sin{Q_2}}{\textstyle  \sqrt{Q_1}\cos{Q_4}}P_2 - \frac{\textstyle \cos{Q_2}\sin{Q_4}}{\textstyle  \sqrt{Q_1}}P_4,	\\
p_2=2\sqrt{Q_1}\sin{Q_2}\cos{Q_4}P_1 +\frac{\textstyle  \cos{Q_2}}{\textstyle  \sqrt{Q_1}\cos{Q_4}}P_2 - \frac{\textstyle \sin{Q_2}\sin{Q_4}}{\textstyle  \sqrt{Q_1}}P_4, \\
p_3=2\sqrt{Q_1}\cos{Q_3}\sin{Q_4}P_1 -\frac{\textstyle  \sin{Q_3}}{\textstyle \sqrt{Q_1}\sin{Q_4}}P_3 + \frac{\textstyle \cos{Q_3}\cos{Q_4} }{\textstyle  \sqrt{Q_1}}P_4,\\
p_4=2\sqrt{Q_1}\sin{Q_3}\sin{Q_4}P_1 +\frac{\textstyle  \cos{Q_3}}{\textstyle \sqrt{Q_1} \sin{Q_4}}P_3 + \frac{\textstyle \sin{Q_3}\cos{Q_4}}{\textstyle  \sqrt{Q_1}}P_4.\\
\end{array}
\eeq
Then in new variables the Hamiltonian will be
$$\overline{{\cal K}} = \frac{1}{8}\Bigl(4Q_1P_1^2+\frac{P_2^2}{Q_1\cos^2Q_4}+ 
\frac{P_3^2}{Q_1\sin^2Q_4}+\frac{P_4^2}{Q_1}\Bigr) + p_0Q_1 + Q_1\overline{V}.
$$ 
In the notations of Stackel theorem we have
$$
c_1=Q_1,\quad c_2=\frac{1}{4Q_1\cos^2Q_4},\quad c_3=\frac{1}{4Q_1\sin^2Q_4},\quad c_4=\frac{1}{4Q_1}.
$$
In this case the solution of (\ref {h14}) will be the matrix
\eq{KK14b}\Phi=\left(\begin{array}{rrrc}
      \frac{\textstyle 1}{\textstyle Q_1}    &  0 &  0 &  0    \\
     0 & 1 &     0 & -\frac{\textstyle 1}{\textstyle \cos^2Q_4}     \\
     0 &  0 & 1 &-\frac{\textstyle 1}{\textstyle \sin^2Q_4}       \\
      \frac{\textstyle 1}{\textstyle 4Q_1^2} & 0 &  0 &   -1    \\
 \end{array}\right).
\eeq
The potential $ \overline {V}$, admitting separation of variables, can be written as
$$
\overline{V}=\frac{1}{Q_1}\Bigl(Q_1\overline{V}_1(Q_1)+\frac{1}{4Q_1\cos^2Q_4}\overline{V}_2(Q_2)+
\frac{1}{4Q_1\sin^2Q_4}\overline{V}_3(Q_3)+\frac{1}{4Q_1}\overline{V}_4(Q_4)\Bigr).
$$
In view of relations 
$$Q_1\cos^2Q_4=q_1^2+q_2^2=\frac{r+x_3}{2},\quad Q_1\sin^2Q_4=q_3^2+q_4^2=\frac{r-x_3}{2},\quad Q_1=|{\bf q}|^2=r,
$$
$$
\tan^2{Q_4}=\frac{q_3^2+q_4^2}{q_1^2+q_2^2}=\frac{1-x_3/r}{1+x_3/r},\quad
Q_2=\arctan\frac{q_2}{q_1},\quad Q_3=\arctan\frac{q_4}{q_3}
$$
following from (\ref {h19}), (\ref {KK14}), and the remarks above, we obtain the required form of potential in $x$-coordinates
 \eq{KK2}
V({\bf x}) =\frac{1}{r}\Bigl[G_1(r)+\frac{2A}{r+x_3}
+\frac{2B}{r-x_3}+\frac{1}{r}G_2\bigl(\frac{x_3}{r}\bigr)\Bigr],
\eeq
where $G_1 $, $G_2 $ are arbitrary smooth functions and $A $, $B $ arbitrary constants.

Now assume that a Hamiltonian (\ref {KK0}) with the potential (\ref {KK2}) is given. 
Applying $L$-transformation (\ref {h19}), we write the new Hamiltonian in $q$-coordinates as
$$
{\cal K} = \frac{1}{8}|{\bf p}|^2 + p_0|{\bf q}|^2 + 
G_1(|{\bf q}|^2)+\frac{A}{q_1^2+q_2^2}+\frac{B}{q_3^2+q_4^2}+
\frac{1}{|{\bf q}|^2}G_2\Bigl(\frac{q_1^2+q_2^2-q_3^2-q_4^2}{|{\bf q}|^2}\Bigr).
$$
Fulfilling canonical transformation (\ref {KK14}), (\ref {KK14a}), we have
$$\overline{{\cal K}} = Q_1\Bigl(\frac{P_1^2}{2}+p_0+\frac{G_1(Q_1)}{Q_1}\Bigr)+
\frac{1}{4Q_1\cos^2Q_4}\Bigl(\frac{P_2^2}{2}+4A\Bigr) +
$$
$$
+\frac{1}{4Q_1\sin^2Q_4}\Bigl(\frac{P_3^2}{2}+4B\Bigr)+\frac{1}{4Q_1}\Bigl(\frac{P_4^2}{2}+4G_2(\cos{2Q_4})\Bigr).
$$
Taking into consideration matrix (\ref {KK14b}), we then obtain  
$$f_1(Q_1)=2\Bigl(\frac{\alpha_1}{Q_1}+\frac{\alpha_4}{4Q_1^2}-p_0-\frac{G_1(Q_1)}{Q_1}\Bigr),
\quad f_2(Q_2)=2(\alpha_2-4A),$$
$$f_3(Q_3)=2(\alpha_3-4B),\quad
f_4(Q_4)=2\Bigl(-\frac{\alpha_2}{\cos^2{Q_4}}-\frac{\alpha_3}{\sin^2{Q_4}}-\alpha_4-4G_2(\cos{2Q_4})\Bigr).$$
The integrals of motion follow from (\ref {h14b}).

Note that using arbitrary $L$-transformations allows to introduce two parameters into the potentials obtained. 
Tthese two parameters are determined by some constant unit vector $ {\bf b} $. For example, instead of 
(\ref {KK2}) one can write 
$$
V({\bf x}) =\frac{1}{r}\Bigl[G_1(r)+\frac{2A}{r+{\bf b}^{T}{\bf x}}
+\frac{2B}{r-{\bf b}^{T}{\bf x}}+\frac{1}{r}G_2\Bigl(\frac{{\bf b}^{T}{\bf x}}{r}\Bigr)\Bigr].
$$
In the next section we show how to perform separation of variables in this case.

\begin{center}
\textbf{3. Integration of the system of equations in a special case}
\end{center}

In this section we perform straightforward integration of a system with potential of the form (\ref {KK1}) having additional parameters. Namely, consider the potential
\eq{K1}
V = V({\bf x}) = -\frac{1}{r}\Bigl(G_1((r+{\bf b}^{T}{\bf x})/2)+G_2((r-{\bf b}^{T}{\bf x})/2)\Bigr),
\eeq
where $G_1$, $G_2$ are some smooth functions, and ${\bf b}=(b_1,b_2,b_3)^{T}$ 
an arbitrary unit vector.  
Note that the vector $ {\bf b} $ provides two parameters in explicit form. 
Having in mind only theoretical investigation (integrability problem), one can take $ {\bf b} $ to be the ort along the $x_1$-axis. On the other hand, from the more practical point of view, introducing vector $ {\bf b} $ gives us additional degree of freedom necessary for applied problems of celestial mechanics. 
In such problems, the axes are usually connected with some special directions (equinox or zenith).
Therefore the presence of the vector $ {\bf b} $ in potential (\ref {K1}) 
allows one to turn the coordinate system at one's will.

As $G_1 $, $G_2 $ one can take, for example, functions of the form
$$
\frac{1}{r}(r+{\bf b}^{T}{\bf x})^k,\quad \frac{1}{r}(r-{\bf b}^{T}{\bf x})^k,
\quad k=1,2,\ldots
$$
 We consider a finite linear combination
\eq{DD2}
V =  -\frac{1}{r}\sum_{k=1}^{N}\Bigl(A_k(r+{\bf b}^{T}{\bf x})^k+
B_k(r-{\bf b}^{T}{\bf x})^k\Bigr).
\eeq
Here $A_k $, $B_k $ are constants. 
Such a potential was considered in  \cite{PSM30}. 
This case leads in general to hyperelliptic integrals. 

For an interested reader here is a problem: find a real perturbing potential which can be approximated by functions of the form (\ref {DD2}).
Note that the combination 
$$
-\frac{B}{4r}(r+{\bf b}^{T}{\bf x})^2+ 
\frac{B}{4r}(r-{\bf b}^{T}{\bf x})^2 = -B{\bf b}^{T}{\bf x}
$$
gives potential corresponding to a constant force. 
Applications of such potential were considered in \cite{Belec,Kunic,demin}.


The canonical equations of motion have the form
\eq{K3}
\begin{array}{ll}
\frac{\textstyle dx_i}{\textstyle dt} = & y_i,\\
\frac{\textstyle dy_i}{\textstyle dt} = &
-\frac{\textstyle\mu}{\textstyle r^3}x_i -\frac{\textstyle x_i}{\textstyle r^3} \Bigl(G_1((r+{\bf b}^{T}{\bf x})/2)+G_2((r-{\bf b}^{T}{\bf x})/2)\Bigr)+\\
 & +\frac{\textstyle 1}{\textstyle 2r} \Bigl(G_1^\prime((r+{\bf b}^{T}{\bf x})/2)(\frac{\textstyle x_i}{\textstyle r}+b_i)+G_2^\prime(r-{\bf b}^{T}{\bf x})/2)(\frac{\textstyle x_i}{\textstyle r}-b_i)\Bigr),
\end{array} 
\eeq
where $i=1,2,3$ and the sign prime indicates the derivative. 
 
This system is the same as the equation of the perturbed two-body problem
$$
\begin{array}{ll}
\ddot{{\bf x}} +  \frac{\textstyle\mu}{\textstyle r^3}{\bf x} =& \frac{\textstyle 1}{\textstyle 2r} \Bigl(G_1^\prime((r+{\bf b}^{T}{\bf x})/2)-
G_2^\prime((r-{\bf b}^{T}{\bf x})/2)\Bigr){\bf b}+\\
  &+\frac{\textstyle 1}{\textstyle 2r^2} \Bigl(G_1^\prime((r+{\bf b}^{T}{\bf x})/2)+G_2^\prime((r-{\bf b}^{T}{\bf x})/2)\Bigr){\bf x}-\\
 &-\frac{\textstyle 1}{\textstyle r^3} \Bigl(G_1((r+{\bf b}^{T}{\bf x})/2)+G_2((r-{\bf b}^{T}{\bf x})/2)\Bigr){\bf x}.
\end{array}
$$
From this one can see that the perturbation is defined by two forces. 
The first force is collinear to the fixed vector $ {\bf b} $, and its module varies in dependence on vector $ {\bf x} $. 
The second force is the central one. 



We are going to show that the system (\ref{K3}) is integrable 
in regular variables found by $L$-transformations.
Transformation (\ref{K7}) contains an arbitrary $L$-matrix. 
A special choice of this matrix allows one to separate the variables in the case of an arbitrary constant 
unitary vector ${\bf b}$.

Consider the term in (\ref{K88}) containing $V_c({\bf q})$. In the new variables this becomes
$$|{\bf q}|^2V_c({\bf q})=-G_1((|{\bf q}|^2+
{\bf q}^{T}(b_1K_1+b_2K_2+b_3K_3)K_4{\bf q})/2)-$$
$$-G_2((|{\bf q}|^2-{\bf q}^{T}(b_1K_1+b_2K_2+b_3K_3)K_4{\bf q})/2).$$
We assume that the $L$-matrix has the first type and $K_4=-{\cal Y}$.
Then
\eq{K10a}|{\bf q}|^2V_c({\bf q})=-G_1((|{\bf q}|^2-C)/2)-G_2((|{\bf q}|^2+C)/2),\eeq
where 
$$ C = {\bf q}^{T}
\Bigl[(b_1a_{11}+b_2a_{12}+b_3a_{13}){\cal U}+$$
$$+(b_1a_{21}+b_2a_{22}+b_3a_{23}){\cal V}+
(b_1a_{31}+b_2a_{32}+b_3a_{33}){\cal W}\Bigr]{\cal Y}{\bf q}.$$
 Let's select parameters $L$-matrixes $a _ {ij} $ from a system (\ref {K11}). Then 
$$ C = {\bf q}^{T}{\cal U}{\cal Y}{\bf q}={\bf q}^{T}\left(\begin{array}{r}
 -q_1\\ -q_2\\ q_3 \\ q_4 \\ \end{array}\right)= -q_1^2-q_2^2 + q_3^2 +q_4^2.$$
Substituting the found value $C $ in (\ref {K10a}), we obtain 
$$|{\bf q}|^2V_c({\bf q})=-G_1(q_1^2+q_2^2)-G_2(q_3^2+q_4^2).$$
It follows that the Hamiltonian (\ref {K88}) is represented in the form of the sum
$${\cal K} = {\cal K}_1 + {\cal K}_2, $$
where
$${\cal K}_1=\frac{1}{8}(p_1^2+p_2^2)+p_0(q_1^2+q_2^2)-G_1(q_1^2+q_2^2),$$
$${\cal K}_2=\frac{1}{8}(p_3^2+p_4^2)+p_0(q_3^2+q_4^2)-G_2(q_3^2+q_4^2).$$
As the value of $p_0 $ is constant, the system (\ref {K8}) splits into two 
independent subsystems
\eq{K12}
\frac{dq_i}{d\tau} = \frac{\partial {\cal K}_1}{\partial p_i},\quad
\frac{dp_i}{d\tau} = -\frac{\partial {\cal K}_1}{\partial q_i},
\quad i=1,2,
\eeq
\eq{K13}
\frac{dq_i}{d\tau} = \frac{\partial {\cal K}_2}{\partial p_i},\quad
\frac{dp_i}{d\tau} = -\frac{\partial {\cal K}_2}{\partial q_i},
\quad i=3,4.
\eeq

We integrate all over again a system (\ref {K12}). In the bipolar coordinates
 Hamiltonian ${\cal K}_1 $, and accordingly the system, have the form 
$$\overline{\cal K}_1=\frac{1}{8}\Bigl(4Q_1P_1^2+\frac{P_2^2}{Q_1}\Bigr)+
p_0Q_1-G_1(Q_1),$$
\eq{K16}
\begin{array}{l}
\frac{\textstyle dQ_1}{\textstyle d\tau} =Q_1P_1,\quad
\frac{\textstyle dQ_2}{\textstyle d\tau} =\frac{\textstyle P_2}{\textstyle 4Q_1}, \\
\frac{\textstyle dP_1}{\textstyle d\tau} =
-\frac{\textstyle 1}{\textstyle 2}P_1^2+\frac{\textstyle P_2^2}{\textstyle 8Q_1^2}-p_0+G_1^\prime(Q_1),\quad
\frac{\textstyle dP_2}{\textstyle d\tau} =0.\\
\end{array}
\eeq

Since the Hamiltonian $\overline{\cal K}_1$   does  not explicitly depend on $\tau$ and $Q_2$, the system (\ref{K16}) has two integrals,
\eq{K17}
\frac{1}{2}Q_1P_1^2+\frac{P_2^2}{8Q_1}+p_0Q_1-G_1(Q_1)=\frac{E_1}{8}.
\eeq
$$
P_2=c_1.$$
Here, $E_1$ and $c_1$  are the constants of integration. Taking these integrals into account, the equation for $P_1$ may be written in the following form
$$
\frac{dP_1}{d\tau} =
\frac{c_1^2}{4Q_1^2}-\frac{E_1}{8Q_1}+G_1^\prime(Q_1)-\frac{G_1(Q_1)}{Q_1}.$$
Eliminating $d\tau$ from equations for $P_1$,  $Q_1$ and integrating the resulting equation, we find
$$
P_1=\frac{\delta_1}{2Q_1}\sqrt{\Phi_1(Q_1)},\quad \delta_1=\pm 1,
$$
where
$$
\Phi_1(Q_1)=-c_1^2+E_1Q_1+c_2Q_1^2+8Q_1G_1(Q_1)
$$
and $c_2$ is integration constant defined by
$$
c_2=4{(P_1^0)^2}+\frac{c_1^2}{{(Q_1^0)^2}}-\frac{E_1}{Q_1^0}-
8\frac{G_1(Q_1^0)}{Q_1^0}.
$$
Due to nonnegativity of $Q_1$, from the first equation of the system (\ref {K16}) it follows that 
$$ \delta_1=\mathop{\rm sign}{Q_1'}.$$
Substituting the derived $P_1$ to the first equation of (\ref{K16}) we find
\eq{K19}
\tau+c_3=2\delta_1\int\limits_{\xi}^{Q_1}{\frac{dQ_1}{\sqrt{\Phi_1(Q_1)}}}.
\eeq
Using the continuity principle, the sign before the integral (\ref {K19}) cannot 
change when $\Phi_1(Q_1)$ is non-zero. 
Therefore, the function $\tau(Q_1)$ in this case behaves monotonically. 
Inverting the integral (\ref {K19}), we obtain $Q_1$ as a function of $\tau$; 
we substitute this function in the second equation of the system (\ref {K16}). Then we get
$$
Q_2=\frac{c_1}{4}\int\limits_{0}^{\tau}{\frac{d\tau}{Q_1(\tau)}}+c_4,\quad c_4=Q_2^0.
$$
Here $c_3$ and $c_4$ are the integration constants. Thus, the values $Q_1$, $Q_2$, $P_1$ are represented as functions of $\tau $. 
If $\Phi_1(Q_1)$ is a polynomial, the integral (\ref {K19}) is, in general, hyperelliptic.

The integration of the system (\ref {K13}) is done similarly. 
We find as a result
$$
Q_4=\frac{c_5}{4}\int\limits_{0}^{\tau}{\frac{d\tau}{Q_3(\tau)}}+c_8,\quad c_8=Q_4^0,
$$
$$
P_3=\frac{\delta_2}{2Q_3}\sqrt{\Phi_2(Q_3)},\quad \delta_2=\mathop{\rm sign}{Q_3'},\quad P_4=c_5,
$$
where
$$
\Phi_2(Q_3)=-c_5^2+E_2Q_3+c_6Q_3^2+8Q_3G_2(Q_3).
$$
Here $c_6$ and $E_2$ are the integration constants defined by the equalities
$$
c_6=4{(P_3^0)^2}+\frac{c_5^2}{{(Q_3^0)^2}}-\frac{E_2}{Q_3^0}-
8\frac{G_2(Q_3^0)}{Q_3^0},$$
\eq{K17a}
\frac{E_2}{8}=\frac{1}{2}Q_3P_3^2+\frac{P_4^2}{8Q_3}+p_0Q_3-G_2(Q_3).\eeq
The function $Q_3(\tau)$ is found by a reversion of the integral
\eq{K19a}
\tau+c_7=2\delta_2\int\limits_{\eta}^{Q_3}{\frac{dQ_3}{\sqrt{\Phi_2(Q_3)}}}.
\eeq
Thus, the values $Q_3$, $Q_4$, and $P_3$ are also determined as functions 
of the variable $\tau$.
The lower limits $\xi$ and $\eta$ in integrals (\ref{K19}) and (\ref{K19a}) are chosen according to the location of $Q_1$ and $Q_3$ with respect to the roots of functions $\Phi_1(Q_1)$ and $\Phi_2(Q_3)$, respectively. 

The formulae of inverse transformation
$$\begin{array}{l}
Q_1=q_1^2+q_2^2,\quad \tan{Q_2}=\frac{\textstyle q_2}{\textstyle q_1},\\
Q_3=q_3^2+q_4^2,\quad \tan{Q_4}=\frac{\textstyle q_4}{\textstyle q_3},\\
P_1=\frac{\textstyle q_1p_1+q_2p_2}{\textstyle 2(q_1^2+q_2^2)},\quad
P_2=-q_2p_1+q_1p_2, \\
P_3=\frac{\textstyle q_3p_3+q_4p_4}{\textstyle 2(q_3^2+q_4^2)},\quad
P_4=-q_4p_3+q_3p_4 \\
\end{array}
$$
allow to define initial values of the variables $Q_i$ and $P_i$ $(i=1,2,3,4)$.

The values of integration constants $c_1$, $c_2$, $c_3$, $c_4$, and $E_1$ are determined by the initial values of $Q_1 $, $Q_2 $, $P_1 $, $P_2$.
These five constant values are connected with each other by the integral (\ref {K17}). 
In the same way, the constant values $c_5$, $c_6$, $c_7$, $c_8$, and $E_2$ are connected by the integral (\ref {K17a}) and are defined by initial values of $Q_3$, $Q_4$, $P_3$, and $P_4$. 
From $p_0 = -H({\bf x}^0,{\bf y}^0)$ we find also relation $E_1+E_2=8\mu$.
One has to add the above relations for $c_2$ and $c_6$ to these connections.
Besides, as the bilinear relation $ {\bf q}^{T} K_4{\bf p} =0 $ 
is the integral of (\ref {K8}), in our case we have
$${\bf q}^{T}(-Y){\bf p}=-q_2p_1+q_1p_2+q_4p_3-q_3p_4=0.$$
Therefore the equality $P_2=P_4$, or equivalently $c_1=c_5$, also holds.

Applying further the first four formulas (\ref {K14}) and (\ref {K15}),
we find $q_i$, $p_i$ $(i=1,2,3,4)$ as functions of $\tau$.
Finally, integrating the two remaining equations of (\ref {K8}), we obtain
$p_0=-H({\bf x}^0,{\bf y}^0)$
and physical time expressed through $\tau$,  
\eq{K19c}
t=q_0=\int\limits_{0}^{\tau}{|{\bf q}|^2d\tau}+c_9= t_1 + t_2,
\eeq
where
$$ t_1=\int\limits_{0}^{\tau}Q_1(\tau)d\tau,\quad 
t_2=\int\limits_{0}^{\tau}Q_3(\tau)d\tau,\quad c_9=0.
$$
Thus, the system (\ref {K8}) is completely integrated and we can, at least locally, find a required trajectory. 
Here it is necessary to note, that if perturbing potentials $G_1$, $G_2$ in (\ref {K3}) are analytic, then, as it is known from a course of the differential equations, the solution of a problem will also be analytic. Let us suppose that the local inversion of integrals (\ref {K19}), (\ref {K19a}) appeared to be a globally determined function. 
In this case we can conclude, by uniqueness of analytic continuation, that this inversion gives not only local, but also global solution of the problem (\ref {K3}). This is the case when functions $G_1$, $G_2$ are polynomials of degree two or three. 
In this case (\ref {K19}) and (\ref {K19a}) are the elliptic integrals, for whose inversion we have at our disposal the well developed technique of elliptic functions; thus, we have found the solution of (\ref {K3}) in explicit form. 

\begin{center}
 \textbf{4. Inversion of the integral in elliptic case }
\end{center}
In this section we consider one case of functions $G_1$ and $G_2$, which reduces to elliptic integrals. Other cases have been studied in \cite {PSM28,PSM29,PSM30,PSM31}.
Take as $G_1$ and $G_2$ the functions
\eq{K20a}G_1=\frac{A_{-1}}{r+{\bf b}^{T}{\bf x}}+A_{1}(r+{\bf b}^{T}{\bf x})+A_2(r+{\bf b}^{T}{\bf x})^2,\eeq
\eq{K20b}G_2=\frac{B_{-1}}{r-{\bf b}^{T}{\bf x}}+B_{1}(r-{\bf b}^{T}{\bf x})+B_2(r-{\bf b}^{T}{\bf x})^2,\eeq
where $A_{-1}$, $A_1$, $A_2$, $B_{-1}$, $B_1$, and $B_2$ are the parameters of the potential. 
Then for the functions $\Phi_1(Q_1)$ and $\Phi_2(Q_3)$ in (\ref {K19}) and (\ref {K19a}) we have the  expressions
$$
\Phi_1(Q_1)=\widehat{c}_1+E_1Q_1+c_2Q_1^2+32A_2Q_1^3,
$$
$$
\Phi_2(Q_3)=\widehat{c}_5+E_2Q_3+c_6Q_3^2+32B_2Q_3^3.
$$
Here $\widehat{c}_1=-c_1^2+4A_{-1}$, $c_2=16A_1-8p_0$, $\widehat{c}_5=-c_5^2+4B_{-1}$, $c_6=16B_1-8p_0$.

Firstly, let us note that the variables $Q_1$ and $Q_3$ are non-negative by definition, 
and that from integrals (\ref {K19}) and (\ref {K19a}) it follows that the ranges of
these variables are determined by the inequalities
\eq{K20}
\Phi_1(Q_1)\ge 0,\quad \Phi_2(Q_3)\ge 0.
\eeq
Let us reverse the integral (\ref {K19}). The number of roots of the polynomial $\Phi_1$ 
and their positions depend on the value of $A_2$. With $A_2 = 0$ the degree of  $\Phi_1(Q_1)$ equals to two. The integral (\ref {K19}) is found in elementary functions, so this case is not being considered here. We distinguish two cases: $A_2 < 0$, $A_2 > 0$. 
Let's note the roots of $\Phi_1(Q_1)$ as $\xi_1$, $\xi_2$, $\xi_3$. 
The cases under consideration will be sequentially numbered by parameter $i_A$. 

\begin{figure}[h,t,p]
\begin{center}
		\includegraphics[width=6.5cm,height=4.5cm  ]{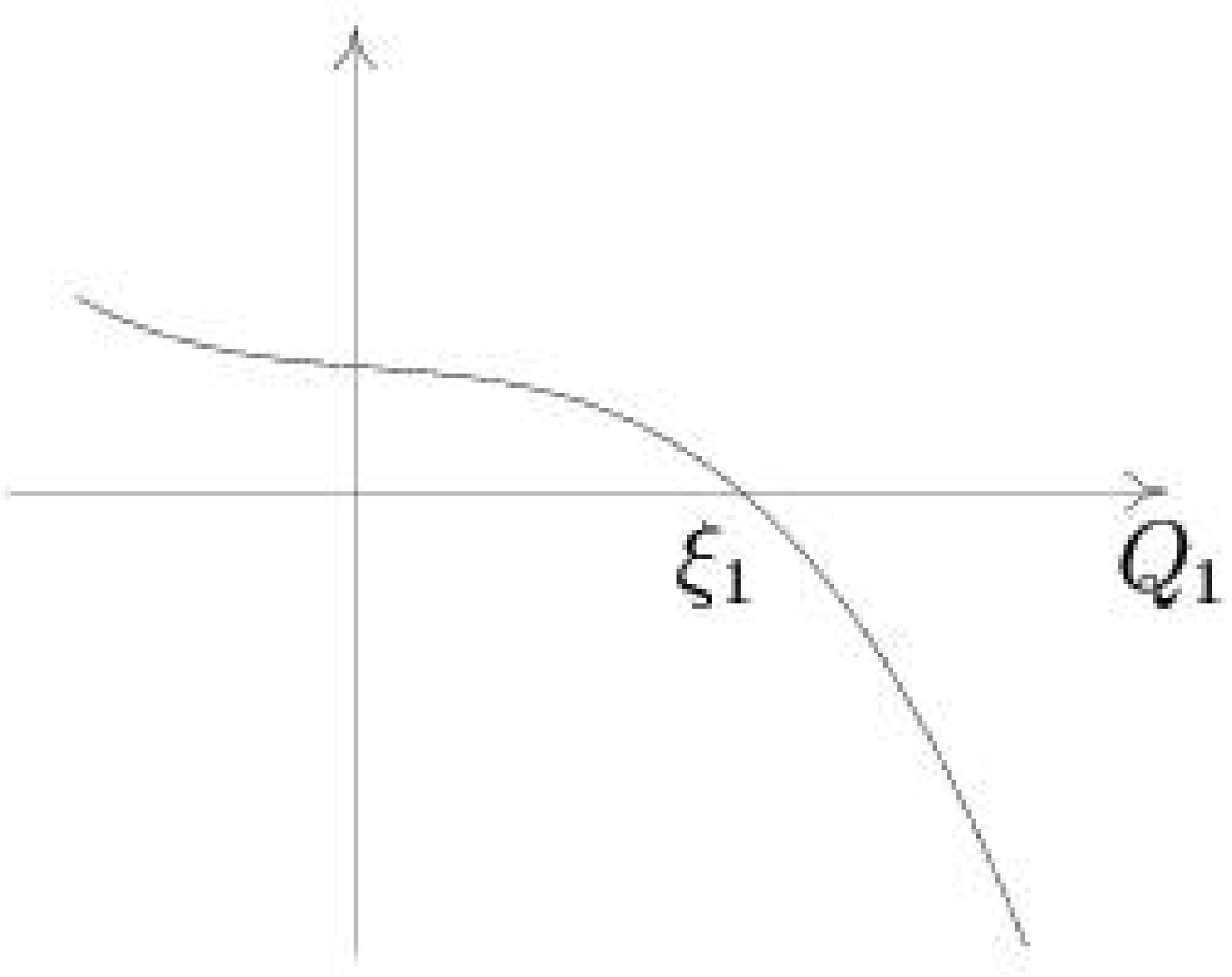}
\end{center}
\caption{The graph of $\Phi_1(Q_1)$. The case $A_2 < 0$. }\label{r1a}
\end{figure}

\begin{figure}[h,t,p]
\begin{center}
		\includegraphics[width=6.5cm,height=4.5cm  ]{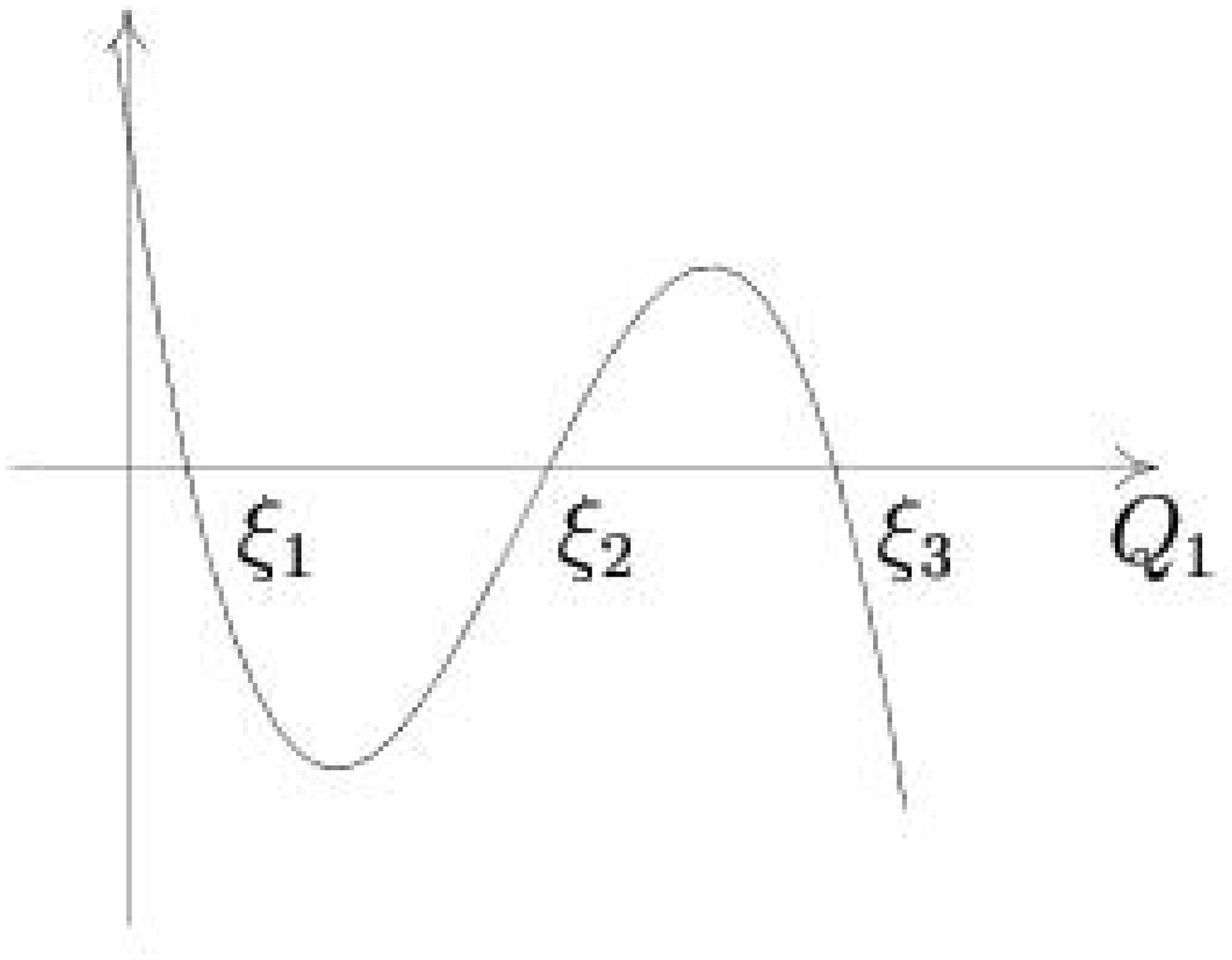}
\end{center}
\caption{The graph of $\Phi_1(Q_1)$. The case $A_2 < 0$. }\label{r1}
\end{figure}

I. Assume that $A_2 < 0$. 
 In this case $\Phi_1(-\infty) > 0$, $\Phi_1(+ \infty) < 0$. 
 The value $\Phi_1 (0) = \widehat {c} _1$ may be both positive and negative. 
 For actual motion there should be at least one positive root. 
 The qualitatively different cases of the graph of $\Phi_1(Q_1)$ are shown in figures \ref {r1a} and \ref {r1}. In the case of three real roots (fig. \ref {r1}), the axis of ordinates 
 goes between $\xi_1$, $\xi_2$ if  $\widehat{c}_1 < 0$, and left with respect to $\xi_1$ or between $\xi_2$, $\xi_3$ if $\widehat{c}_1 > 0$.  

The case $i_A=1$.
Suppose that $\Phi_1$ has one real root $\xi_1$, and that $Q_1^0\in (0, \xi_1)$ (fig. \ref{r1a}).
Let's write the integral (\ref {K19}) in the form
$$
\tau+c_3=\frac{\delta_1}{2\sqrt{-2A_2}}
\int\limits_{\xi_1}^{Q_1}{\frac{dz}{\sqrt{(\xi_1-z)(z^2+bz+c)}}},
$$
where the square trinomial $z^2+bz+c $ has no real roots and is positive for all $z$, and
\eq{K22f}
b=\xi_1+\frac{c_2}{32A_2},\quad c=b\xi_1+\frac{E_1}{32A_2}\quad (c>0).
\eeq
Apply in the integral the substitution
$$ z=\xi_1-a \frac{1-\cos\varphi}{1+\cos\varphi},\quad
a=\sqrt{\xi_1^2+b\xi_1+c}
$$
and put the notations 
$$ 
\varphi_1=2\arctan\sqrt{\frac{\xi_1-Q_1}{a}},\quad k_1^2=\frac12\Bigl(1+\frac{\xi_1+b/2}{a}\Bigr),\quad l_1=2\sqrt{-2aA_2}.
$$

Then we derive
\eq{K22b}
\tau+c_3=-\frac{\delta_1}{l_1}
\int\limits_{0}^{\varphi_1}{\frac{d\varphi}{\sqrt{1-k_1^2\sin^2{\varphi}}}}.
\eeq
Putting here $ \tau=0$, we find an integration constant $c_3$: 
$$
c_3=-\frac{\mathop{\rm sign}{P_1^0}}{l_1}
\int\limits_{0}^{\varphi_1^0}{\frac{d\varphi}{\sqrt{1-k_1^2\sin^2{\varphi}}}},
\quad \varphi_1^0=2\arctan\sqrt{\frac{\xi_1-Q_1^0}{a}}. $$

Check that $k_1^2 < 1 $. As $z^2+bz+c $ has no real roots, we have $b^2-4c < 0$. 
Therefore, 
$$\Bigl(\xi_1+\frac{b}{2}\Bigr)^2 < \xi_1^2+b\xi_1+c = a^2
\Rightarrow \Bigl|\frac{\xi_1+b/2}{a}\Bigr|<1.$$
Hence,  $|k_1| < 1$.
Reversing the integral (\ref {K22b}) derived above, we come to the function
$$
Q_1=\xi_1+a-\frac{2a}{1+{\rm cn}(l_1(\tau+c_3);k_1)}.
$$
It is easily to see that for $Q_1\in (0,\xi_1)$ the denominator ${\rm cn}(u)+1 \neq 0$. Calculating the derivative of $Q_1$, we get
$\delta_1=-\mathop{\rm sign}{\mathop{\rm sn}{(l_1(\tau+c_3);k_1)}}$.
For the variable $Q_2$ we find
$$
Q_2=\frac{c_1\tau}{4(\xi_1+a)}+\frac{ac_1}{2l_1(\xi^2_1-a^2)}
\Biggl[\int\limits_{0}^{l_1(\tau+c_3)}{\frac{d u}{1+n_1\mathop{\rm cn}(u;k_1)}}
-\int\limits_{0}^{l_1c_3}{\frac{d u}{1+n_1\mathop{\rm cn}(u;k_1)}}\Biggr]+Q_2^0,
$$
$$
n_1=1+\frac{2a}{\xi_1-a}.
$$
Note that  
$$
\frac{n_1^2}{n_1^2-1}-k_1^2=\frac{c}{4a\xi_1}>0.
$$
Therefore for calculating the integral of
$(1+n_1\mathop{\rm cn}(u;k_1))^{-1}$ we apply the formula (341.03) \cite{Byrd}
\eq{K21h}
\int\limits_{0}^{u}{\frac{d u}{1+n\mathop{\rm cn}(u;k)}}=
\frac{1}{1-n^2}\Bigl[\Pi\Bigl(u,\frac{n^2}{n^2-1};k \Bigr)-n g_1\Bigr],\quad n^2 \ne 1,
\eeq
with
\eq{K21f}
g_1(u)=\frac{1}{2}\sqrt{\frac{n^2-1}{k^2+k'^2n^2}}
\ln\Biggl|\frac{\sqrt{k^2+k'^2n^2}\mathop{\rm sn}(u;k)+\sqrt{n^2-1}\mathop{\rm dn}(u;k)}
{\sqrt{k^2+k'^2n^2}\mathop{\rm sn}(u;k)-\sqrt{n^2-1}\mathop{\rm dn}(u;k)}\Biggr|,\quad k'^2= 1-k^2.
\eeq
Here we note the elliptic integral of the third kind as
$$
\Pi(u,n;k)\equiv\int\limits_{0}^{u}{\frac{d v}{1-n{\rm sn}^2(v;k)}}.
$$
For $t_1$ we have
$$
t_1 =(\xi_1+a)\tau - \frac{2a}{l_1}\Biggl[
\int\limits_{0}^{l_1(\tau+c_3)}\frac{d u}{1+{\rm cn}(u;k_1)}-
\int\limits_{0}^{l_1c_3}\frac{d u}{1+{\rm cn}(u;k_1)}  \Biggr]. $$
The integral of $(1+\mathop{\rm cn}(u;k_1))^{-1}$ is calculated by the formula 
(341.53) \cite {Byrd}  
\eq{K22c}
\int\limits_{0}^{u}\frac{d v}{{1\pm \rm cn}(v;k)}=
u- E(u)\pm \frac{\mathop{\rm dn}(u;k) \mathop{\rm sn} (u;k)}{1 \pm\mathop{\rm cn}(u;k)}, \eeq
where $E(u)=E(\varphi;k)$ is incomplete elliptic integral of the second kind  
$(\varphi=\mathop{\rm am} u)$.

The case $i_A=2$.
Suppose that $\Phi_1(Q_1)$ has three real roots $0 < \xi_1 < \xi_2 < \xi_3 $, and 
$Q_1^0\in (0,\xi_1)$. Let's write (\ref {K19}) as
$$
\tau+c_3=\frac{\delta_1}{2\sqrt{-2A_2}}
\int\limits_{Q_1}^{\xi_1}{\frac{dz}{\sqrt{(\xi_1-z)(\xi_2-z)(\xi_3-z)}}}.
$$
Making the substitution 
$ \varphi=\arcsin{\sqrt{(\xi_1-z)/(\xi_2-z)}}$ 
and reversing the resulting integral, we find
$$
Q_1=\xi_2-\frac{(\xi_2-\xi_1)}{{\rm cn}^2{(l_1(\tau+c_3);k_1)}},
$$
where
$$
k_1=\sqrt{\frac{\xi_3-\xi_2}{\xi_3-\xi_1}},\quad
l_1=\sqrt{-2A_2(\xi_3-\xi_1)},
$$
$$
c_3=\frac{\mathop{\rm sign}{P_1^0}}{l_1}
\int\limits_{0}^{\varphi_1^0}\frac{d\varphi}{\sqrt{1-k_1^2\sin^2{\varphi}}},
\quad \varphi_1^0=\arcsin{\sqrt{\frac{\xi_1-Q_1^0}{\xi_2-Q_1^0}}}. $$
Now we calculate $\delta_1$. We differentiate $Q_1$ and use the formula of double argument for  elliptic sine. We have
$$
Q_1'=2l_1(\xi_3-\xi_2){\rm cn}^{-3}(u;k_1)(-1)\mathop{\rm sn}(u;k_1){\rm dn}(u;k_1)=$$
$$
=-l_1(\xi_3-\xi_2){\rm cn}^{-4}(u;k_1)(1-k_1^2{\rm sn}^4(u;k_1)){\rm sn}{(2u;k_1)},
$$
where the notation $u=l_1(\tau+c_3)$ is introduced for brevity. Therefore,
 $$\delta_1=\mathop{\rm sign}{Q_1'}=-\mathop{\rm sign}{\mathop{\rm sn}{(2l_1(\tau+c_3);k_1)}}.$$
Now we find $Q_2$  
$$
Q_2=\frac{c_1\tau}{4\xi_2}+ \frac{c_1(\xi_2-\xi_1)}{4l_1\xi_1\xi_2}
\Bigl[\Pi(l_1(\tau+c_3),n_1;k_1)-\Pi(l_1c_3,n_1;k_1)\Bigr]
+Q_2^0,\quad n_1=\frac{\xi_2}{\xi_1}.
$$
For the value of physical time, corresponding to the variable $Q_1$, we have
$$
t_1 =\xi_2\tau + \frac{\xi_1-\xi_2}{l_1}\Biggl[
\int\limits_{0}^{l_1(\tau+c_3)}\frac{d u}{{\rm cn}^2(u;k_1)}-
\int\limits_{0}^{l_1c_3}\frac{d u}{{\rm cn}^2(u;k_1)}  \Biggr], $$
where the integral from ${\rm cn}^{-2}(u;k_1)$ is calculated by the formula (313.02) \cite{Byrd} 
$$\int\limits_{0}^{u}\frac{d v}{{\rm cn}^2(v;k)}=
\frac{1}{1-k^2}\Bigl((1-k^2)u- E(u)+\frac{\mathop{\rm dn}(u;k) \mathop{\rm sn} (u;k)}{\mathop{\rm cn}(u;k)}\Bigr). $$

The case $i_A=3$. 
The polynomial $\Phi_1(Q_1)$ has three real roots $\xi_1 < \xi_2 < \xi_3 $,
and $Q_1^0\in (\max\{0,\xi_2\},\xi_3)$. 
Write (\ref {K19}) as
$$
\tau+c_3=\frac{\delta_1}{2\sqrt{-2A_2}}
\int\limits_{\xi_3}^{Q_1}{\frac{dz}{\sqrt{(z-\xi_1)(z-\xi_2)(\xi_3-z)}}}.
$$
The reduction of this integral to the standard form (\ref {K22b}) is carried out by the substitution $ \varphi=\arcsin{\sqrt{(\xi_3-z)/(\xi_3-\xi_2)}}$. 
The result of reversion can be presented in the form
$$
Q_1=\xi_3+(\xi_2-\xi_3){\rm sn}^2{(l_1(\tau+c_3);k_1)},
$$
where the following notations are used
$$
k_1=\sqrt{\frac{\xi_3-\xi_2}{\xi_3-\xi_1}},\quad
l_1=\sqrt{-2A_2(\xi_3-\xi_1)},
$$
$$
c_3=-\frac{\mathop{\rm sign}{P_1^0}}{l_1}
\int\limits_{0}^{\varphi_1^0}\frac{d\varphi}{\sqrt{1-k_1^2\sin^2{\varphi}}},
\quad \varphi_1^0=\arcsin{\sqrt{\frac{\xi_3-Q_1^0}{\xi_3-\xi_2}}}. $$
For $ \delta_1 $ we find
 $\delta_1=-\mathop{\rm sign}{\mathop{\rm sn}{(2l_1(\tau+c_3);k_1)}}$.
Substitute $Q_1$ in the formulae for $Q_2$ and $t_1 $. We find
$$
Q_2=\frac{c_1}{4l_1\xi_3}\Bigl[\Pi(l_1(\tau+c_3),n_1;k_1)-\Pi(l_1c_3,n_1;k_1)\Bigr]+Q_2^0,
$$
$$
t_1 =\xi_3\tau + \frac{\xi_2-\xi_3}{l_1}\Biggl[
\int\limits_{0}^{l_1(\tau+c_3)}{\rm sn}^2(u;k_1)d u-
\int\limits_{0}^{l_1c_3}{\rm sn}^2(u;k_1)d u\Biggr]. $$
The integral from squared elliptic sine is calculated by the formula \cite{Byrd}
$$
\int\limits_{0}^{u}{{\rm sn}^2(v;k)d v}=\frac{1}{k^2}(u-E(u)).
$$

\begin{figure}[h,t,p]
\begin{center}
		\includegraphics[width=6.cm,height=4.cm  ]{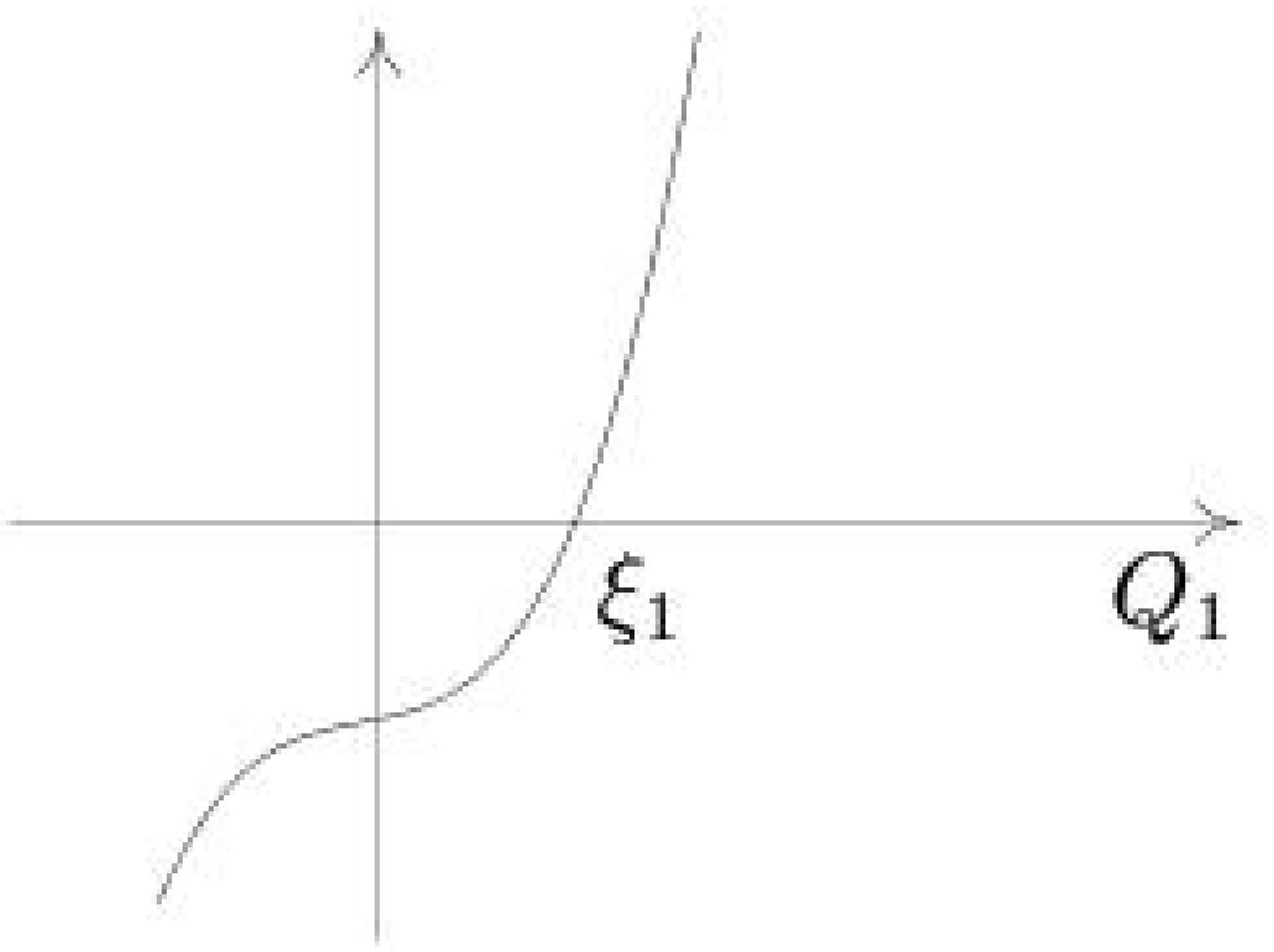}
\end{center}
\caption{The graph $\Phi_1(Q_1)$. Case $A_2 > 0$. }\label{r2}
\end{figure}

\begin{figure}[h,t,p]
\begin{center}
		\includegraphics[width=6.5cm,height=5.cm  ]{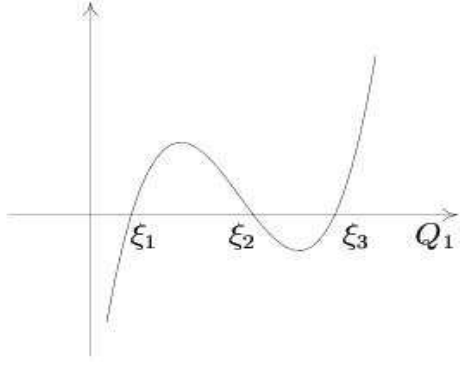}
\end{center}
\caption{The graph $\Phi_1(Q_1)$. Case $A_2 > 0$. }\label{r3}
\end{figure}

II. Assume further that $A_2 > 0 $.
Now we have $\Phi_1(-\infty) < 0$, $\Phi_1(+\infty) > 0$, and $\Phi_1(0)=\widehat{c}_1$.
The qualitatively different cases of the graph $\Phi_1(Q_1)$ are shown in figures \ref{r2} and \ref{r3}.

The case $i_A=4$. The polynomial $ \Phi_1(Q_1) $ has one
real root $\xi_1$ and, accordingly, $Q_1(0)> \max\{0,\xi_1\}$.
The graph of $ \Phi_1(Q_1)$ in this case is shown in fig. \ref {r2}. 
Write the integral (\ref {K19}) as
$$
\tau+c_3=\frac{\delta_1}{2\sqrt{2A_2}}
\int\limits_{\xi_1}^{Q_1}{\frac{dz}{\sqrt{(z-\xi_1)(z^2+bz+c)}}},
$$
where the square trinomial $z^2+bz+c > 0 $ for all $z$.
The coefficients $b$ and $c$ are found by the formulae (\ref {K22f}). 
Applying the substitution 
$$ z=\xi_1+a \frac{1-\cos\varphi}{1+\cos\varphi},\quad
a=\sqrt{\xi_1^2+b\xi_1+c}
$$
and reversing the resulting integral, we come to the function
$$
Q_1=\xi_1-a+\frac{2a}{1+{\mathop{\rm cn}}(l_1(\tau+c_3);k_1)},
$$
where 
$$ 
k_1^2=\frac12\Bigl(1-\frac{\xi_1+b/2}{a}\Bigr),\quad l_1=2\sqrt{2aA_2}.
$$
$$
c_3=\frac{\mathop{\rm sign}{P_1^0}}{l_1}
\int\limits_{0}^{\varphi_1^0}{\frac{d\varphi}{\sqrt{1-k_1^2\sin^2{\varphi}}}},
\quad \varphi_1^0=2\arctan\sqrt{\frac{Q_1^0-\xi_1}{a}}. $$
As above, one can show that $k_1^2 < 1$.
The resulting function $Q_1(\tau)$ is unbounded, as it has an infinite
number of poles on real straight line, which are found by the formula
$$
\tau =\frac{4m+2}{l_1}K(k_1)-c_3,\quad m \in \textbf{Z}.
$$
Further we find that $\delta_1=\mathop{\rm sign}{\mathop{\rm sn}{(l_1(\tau+c_3);k_1)}}$.
For variable $Q_2 $ we have
$$
Q_2=\frac{c_1\tau}{4(\xi_1-a)}-\frac{ac_1}{2l_1(\xi^2_1-a^2)}
\Biggl[\int\limits_{0}^{l_1(\tau+c_3)}{\frac{d u}{1+n_1\mathop{\rm cn}(u;k_1)}}
-\int\limits_{0}^{l_1c_3}{\frac{d u}{1+n_1\mathop{\rm cn}(u;k_1)}}\Biggr]+Q_2^0,
$$
$$
n_1=1-\frac{2a}{\xi_1+a}.
$$
Note that 
$$
\frac{n_1^2}{n_1^2-1}=1+\frac{1}{n_1^2-1}=-\frac{(\xi_1-a)^2}{4a\xi_1}<0<k_1^2.
$$
Therefore for calculating the integral from the function $(1+n_1\mathop{\rm cn}(u;k_1))^{-1}$ the formula (\ref{K21h}) is to be applied  with 
$$g_1=\sqrt{\frac{1-n^2}{k^2+k'^2n^2}}
\arctan\Biggl[\sqrt{\frac{k^2+k'^2n^2}{1-n^2}}\frac{\mathop{\rm sn}(u;k)}{\mathop{\rm dn}(u;k)}\Biggr],\quad k'^2= 1-k^2.
$$
If $ \xi_1=0 $ then $n_1 =-1 $, and for calculating $Q_2$ the formula (\ref {K22c}) should be used. For $t_1 $ we have
$$
t_1 =(\xi_1-a)\tau + \frac{2a}{l_1}\Biggl[
\int\limits_{0}^{l_1(\tau+c_3)}\frac{d u}{1+{\rm cn}(u;k_1)}-
\int\limits_{0}^{l_1c_3}\frac{d u}{1+{\rm cn}(u;k_1)}  \Biggr]. $$

 
Suppose that $\Phi_1$ has three real roots $\xi_1 < \xi_2 < \xi_3$. 
The graph of the function $\Phi_1(Q_1)$ in this case is given in fig. \ref {r3}.
This case also splits into two subcases: $\xi_1 < Q_1^0 < \xi_2$ and $\xi_3 < Q_1^0$.

The case $i_A=5$. Suppose that $Q_1^0 \in (\max\{0,\xi_1\}, \xi_2)$. 
We write (\ref{K19}) as
$$
\tau+c_3=\frac{\delta_1}{2\sqrt{2A_2}}
\int\limits_{\xi_1}^{Q_1}\frac{dz}{\sqrt{(z-\xi_1)(z-\xi_2)(z-\xi_3)}}.
$$
We apply to this integral the substitution
$ \varphi=\arcsin{\sqrt{(z-\xi_1)/(\xi_2-\xi_1)}}$
and use the notations
$$
k_1=\sqrt{\frac{\xi_2-\xi_1}{\xi_3-\xi_1}},\quad
l_1=\sqrt{2A_2(\xi_3-\xi_1)}.
$$
Then our integral has the standard form
\eq{K21}
\tau+c_3=\frac{\delta_1}{l_1}
\int\limits_{0}^{\varphi_1}\frac{d\varphi}{\sqrt{1-k_1^2\sin^2{\varphi}}}.
\eeq
Reversing (\ref{K21}) and using the inverse substitution, we find the
required function
$$
Q_1=\xi_1+(\xi_2-\xi_1){\rm sn}^2{(l_1(\tau+c_3);k_1)},
$$
where
$$
c_3=\frac{\mathop{\rm sign}{P_1^0}}{l_1}
\int\limits_{0}^{\varphi_1^0}\frac{d\varphi}{\sqrt{1-k_1^2\sin^2{\varphi}}},
\quad \varphi_1^0=\arcsin{\sqrt{\frac{Q_1^0-\xi_1}{\xi_2-\xi_1}}}. $$
As above, one can show that 
$\delta_1=\mathop{\rm sign}{\mathop{\rm sn}{(2l_1(\tau+c_3);k_1)}}$.
For $Q_2 $ we find
$$
Q_2=\frac{c_1}{4l_1\xi_1}\Bigl[\Pi(l_1(\tau+c_3),n_1;k_1)-\Pi(l_1c_3,n_1;k_1)\Bigr]
+Q_2^0,\quad n_1=1-\frac{\xi_2}{\xi_1}.
$$
For the first summand of physical time $t$ in (\ref{K19c}) we have 
$$
t_1 =\xi_1\tau + \frac{\xi_2-\xi_1}{l_1}\Biggl[
\int\limits_{0}^{l_1(\tau+c_3)}{{\rm sn}^2(u;k_1)d u}-
\int\limits_{0}^{l_1c_3}{{\rm sn}^2(u;k_1)d u}\Biggr]. $$

The case $i_A=6$. Suppose $Q_1^0 \in (\max\{0,\xi_3\}, \infty)$.
The integral (\ref{K19}) has the form
$$
\tau+c_3=\frac{\delta_1}{2\sqrt{2A_2}}
\int\limits_{\xi_3}^{Q_1}\frac{dz}{\sqrt{(z-\xi_1)(z-\xi_2)(z-\xi_3)}}.
$$
Transforming this integral to the standard form (\ref{K21}) is made using the substitution 
$ \varphi=\arcsin{\sqrt{(z-\xi_3)/(z-\xi_2)}}$.
The resulting reversion of the integral in this case is following
$$
Q_1=\xi_2+\frac{\xi_3-\xi_2}{{\rm cn}^2(l_1(\tau+c_3);k_1)},
$$
where
$$
k_1=\sqrt{\frac{\xi_2-\xi_1}{\xi_3-\xi_1}},\quad
l_1=\sqrt{2A_2(\xi_3-\xi_1)},
$$
$$
c_3=\frac{\mathop{\rm sign}{P_1^0}}{l_1}
\int\limits_{0}^{\varphi_1^0}{\frac{d\varphi}{\sqrt{1-k_1^2\sin^2{\varphi}}}},
\quad \varphi_1^0=\arcsin{\sqrt{\frac{Q_1^0-\xi_3}{Q_1^0-\xi_2}}}. $$
Now the function $Q_1(\tau)$ has an infinite number of poles of
the second order, hence it is unbounded. The poles are found by the formula
$$
\tau =\frac{2m+1}{l_1}K(k_1)-c_3,\quad m \in \textbf{Z}.
$$
Further we find the values $\delta_1$, $Q_2$, $t_1$
 $$\delta_1=\mathop{\rm sign}{\mathop{\rm sn}{(2l_1(\tau+c_3);k_1)}},$$
$$
Q_2=\frac{c_1\tau}{4\xi_2}+ \frac{c_1(\xi_2-\xi_3)}{4l_1\xi_2\xi_3}
\Bigl[\Pi(l_1(\tau+c_3),n_1;k_1)-\Pi(l_1c_3,n_1;k_1)\Bigr]
+Q_2^0,\quad n_1=\frac{\xi_2}{\xi_3},
$$
$$
t_1 =\xi_2\tau + \frac{\xi_3-\xi_2}{l_1}\Biggl[
\int\limits_{0}^{l_1(\tau+c_3)}\frac{d u}{{\rm cn}^2(u;k_1)}-
\int\limits_{0}^{l_1c_3}\frac{d u}{{\rm cn}^2(u;k_1)}  \Biggr]. $$

An inversion of the integral (\ref {K19a}) is fulfilled in a similar way.
This integral and the function $\Phi_2$ differ only by notations
from the integral (\ref {K19}) and the function $\Phi_1 $. Therefore, after some evident
renaming, we find the expressions for $Q_3 $, $Q_4 $, and $t_2 $. 
We number these cases sequentially by the parameter $i_B$. Then we have:

$i_B=1\quad \Leftrightarrow\quad B_2<0,\>\eta_2,\eta_3\in {\bf C},\> 0<Q_3^0<\eta_1\quad (Q_3 \textrm{ is bounded}).$

$i_B=2\quad \Leftrightarrow\quad B_2<0,\> 0< Q_3^0<\eta_1<\eta_2<\eta_3\quad 
(Q_3 \textrm{ is bounded}).$

$i_B=3\quad \Leftrightarrow\quad B_2<0,\>  \eta_1<\eta_2<Q_3^0<\eta_3\quad 
(Q_3 \textrm{  is bounded}).$

$i_B=4\quad \Leftrightarrow\quad B_2>0,\> \eta_2,\eta_3\in \textbf{C},\>  \eta_1<Q_3^0\quad (Q_3 \textrm{ is unbounded}).$

$i_B=5\quad \Leftrightarrow\quad B_2>0,\> \eta_1<Q_3^0<\eta_2<\eta_3\quad 
(Q_3 \textrm{ is  bounded}).$

$i_B=6\quad \Leftrightarrow\quad B_2>0,\> \eta_1<\eta_2<\eta_3<Q_3^0\quad 
(Q_3 \textrm{  is unbounded}).$


The study above yield the following theorem.

\textbf{THEOREM 3.} 
\textit{The motion of the particle is bounded if and only if at the initial moment both variables $Q_1$ and~$Q_3 $ are 
restricted on the right by the roots of the polynomials $\Phi_1$ and~$\Phi_2$, correspondingly.}

Now we give a definition of retaining potential, introduced in~\cite {PSM29}.

DEFINITION. 
A potential is named as \textit{retaining}, if for arbitrary initial conditions the motion of a particle in a perturbed field corresponding to this potential is bounded.

Thus, potential (\ref{K1}), where $G_1$, $G_2$ are defined by the formulae (\ref{K20a}), (\ref {K20b}) for $A_2 < 0 $ and $B_2 < 0 $, is retaining.
Generally, the formulae (\ref {K19}), (\ref {K19a}) are not elliptic integrals, and we cannot present a solution in explicit form. Nevertheless, the above-stated qualitative result remains true~\cite{PSM30}.

\begin{center}
 \textbf{5. Numerical examples and analysis of motions }
\end{center}
 
In the examples below we consider the motion of a particle in perturbed 
gravitational field of a planet with spherical density distribution, 
whose gravitational parameter is taken to be $\mu=398601.3~\textrm{km}^3/s^2$. 
The perturbing force is defined by the potential (\ref{K1}), with $G_1$ and $G_2$ calculated by the formulae (\ref{K20a}) and (\ref {K20b}). For convenience (to have no fractions), a  dimensionless direction vector $ \widehat{\bf b}$ for the constant force is used. While doing calculations, this direction vector is assumed to be normalized.
The dimensions of parameters
$A_{-1} $ and $B_{-1} $ are $ [\textrm{km}^4/s^2] $, 
$A_1 $ and $B_1 $ are $ [\textrm {km}^2/s^2] $, 
$A_2 $ and $B_2 $ are $ [\textrm {km}/s^2] $. 
Calculations and construction of orbits were performed using the Maple system with 32 digits.
In each example, for convenience of its analysis, the values of circular and parabolic velocities $v_{cir} $, $v_{par}$ of Keplerian motion are given. 
The perturbations being considered are great, they are non-typical for the Earth's satellites. 
For this reason, we do not give Keplerian elements of osculating orbits for the corresponding initial values.
The initial position of a particle is marked by a point on the corresponding figure.
      
\prim Initial values of coordinates and velocities of a particle:
$$x_1=8200~\textrm{km},\quad x_2=0~\textrm{km}, \quad x_3=6000~\textrm{km},$$
$$\dot{x}_2 = 8.6~\textrm{km/s},\quad \dot{x}_1= \dot{x}_3=0~\textrm{km/s}
\quad (v_{cir}\approx 6.26~\textrm{km/s},\quad v_{par}\approx 8.86~\textrm{km/s}).$$
In an unperturbed case these values define an elliptic  motion.

Parameters of potential are as follows:
$$A_{-1}=0.004~\textrm{km}^4/s^2,\quad A_1=0.06~\textrm{km}^2/s^2,\quad A_2=0.2\cdot 10^{-7}~\textrm{km}/s^2,$$
$$B_{-1}=0.0001~\textrm{km}^4/s^2,\quad B_1=0.008~\textrm{km}^2/s^2,\quad B_2=-0.3\cdot 10^{-4}~\textrm{km}/s^2. $$
Coordinates of direction vector are $\widehat{{\bf b}}=(-1, 2,1)^{T}$. 
In the case under consideration the roots of polynomials $\Phi_1$ and $\Phi_2$ are
$$\xi_1\approx 1478,\quad \xi_2\approx 115346,\quad Q_1^0\approx 4631 \quad
\Rightarrow Q_1^0 \in (\xi_1, \xi_2),$$ 
$$\eta_2\approx 1707, \quad\eta_3\approx 31031,\quad Q_3^0\approx 5529 \quad
\Rightarrow Q_3^0 \in (\eta_2, \eta_3).$$
Therefore, the motion is bounded. This is the case  $i_A=5$, $i_B=3$.

The coordinates and velocities have been calculated during a time range, 
corresponding to two revolutions of the particle around the attracting centre 
without perturbations, that is $ \tau\in [0,2T] $, where $T $ is calculated by the formula 
\eq{K50}
T = \pi\sqrt{-\frac{2}{h_k}},\quad h_k = \frac{|\dot{{\bf x}}^0|^2}{2} -\frac{\mu}{|{\bf x}^0|}.
\eeq
Here $h_k $ is the Keplerian energy. 
Let's remind that $L$-transformation doubles the angles at the origin of coordinates.

\begin{figure}[h,t,p]
	\begin{center}
		\includegraphics[width=8cm,height=4.5cm  ]{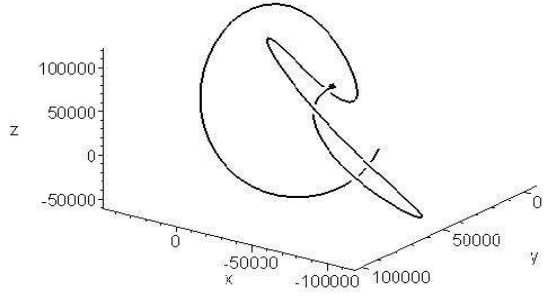}
	\end{center}
		\caption{The case $i_A=5$, $i_B=3$.}	
    \label{sm3_53}
\end{figure}

Note that in this example the potential is not retaining. Nevertheless, the motion appears to be bounded. 
The trajectory of the particle is shown in fig. \ref{sm3_53}.

\prim Initial values of coordinates and velocities of a particle:
$$x_1=8200~\textrm{km},\quad x_2=0~\textrm{km}, \quad x_3=6000~\textrm{km},$$
$$\dot{x}_2 = 9.9~\textrm{km/s},\quad \dot{x}_1= \dot{x}_3=0~\textrm{km/s}
\quad (v_{cir}\approx 6.26~\textrm{km/s},\quad v_{par}\approx 8.86~\textrm{km/s}).$$
In unperturbed case the motion belongs to hyperbolic type.

\begin{figure}[h,t,p]
	\begin{center}
		\includegraphics[width=7cm,height=5cm  ]{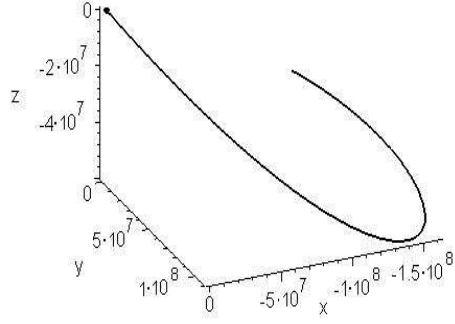}
	\end{center}
		\caption{The case $i_A=3$, $i_B=3$. The hyperbolic type is in unperturbed case.}
		\label{sm2_33}	
\end{figure}

Parameters of potential are as follows: 
$$A_{-1}=0.004~\textrm{km}^4/s^2,\quad A_1=0.006~\textrm{km}^2/s^2,\quad A_2=-0.2\cdot 10^{-7}~\textrm{km}/s^2,$$
$$B_{-1}=0.0001~\textrm{km}^4/s^2,\quad B_1=0.008~\textrm{km}^2/s^2,\quad B_2=-0.3\cdot 10^{-7}~\textrm{km}/s^2. $$
As $A_2 $ and $B_2 $ are negative we have a retaining potential. Coordinates of direction vector are $\widehat{{\bf b}}=(1, 2,-1)^{T}$. 
The roots of polynomials:
$$\xi_2\approx 2126,\quad \xi_3\approx 122192633,\quad Q_1^0\approx  5529\quad
\Rightarrow Q_1^0 \in (\xi_2, \xi_3),$$  
$$\eta_2\approx 1699, \quad\eta_3\approx 81506371,\quad Q_3^0\approx 4631 \quad
\Rightarrow Q_3^0 \in (\eta_2, \eta_3),$$
Therefore, the motion is bounded. This is the case $i_A=3$, $i_B=3$.
The integration is carried out during the time range corresponding approximately to $t=1759.74$ days. The particle trajectory is shown in fig. \ref{sm2_33}.

\prim Initial values of coordinates and velocities of a particle are as follows:
$$x_1=6000~\textrm{km},\quad x_2=0~\textrm{km}, \quad x_3=-8000~\textrm{km},$$
$$\dot{x}_2 = 7.9~\textrm{km/s},\quad \dot{x}_1= \dot{x}_3=0~\textrm{km/s}
\quad (v_{cir}\approx 6.31~\textrm{km/s},\quad v_{par}\approx 8.93~\textrm{km/s}).$$
In an unperturbed case these values define an elliptic  motion. 

Parameters of a potential are as follows:
$$A_{-1}=0.04~\textrm{km}^4/s^2,\quad A_1=0.03~\textrm{km}^2/s^2,\quad A_2=-0.2\cdot 10^{-5}~\textrm{km}/s^2,$$
$$B_{-1}=0.1\cdot 10^{-4}~\textrm{km}^4/s^2,\quad B_1=-0.0003~\textrm{km}^2/s^2,\quad B_2=0.3\cdot 10^{-4}~\textrm{km}/s^2. $$
Here the potential is not retaining. Coordinates of direction vector are $\widehat{{\bf b}}=(1, 1,1)^{T}$.
The roots of polynomials are as follows: 
$$\xi_2\approx 2686,\quad \xi_3\approx 20699,\quad Q_1^0\approx 4423 \quad
\Rightarrow Q_1^0 \in (\xi_2, \xi_3),$$ 
$$\eta_1\approx 3256, \quad\eta_2, \eta_3 \in {\bf C},\quad Q_3^0\approx 5577 \quad
\Rightarrow Q_3^0 > \eta_1.$$
Therefore, the motion is unbounded. This is the case  $i_A=3$, $i_B=4$.
\begin{figure}[h,t,p]
	\begin{center}
		\includegraphics[width=8cm,height=4.5cm  ]{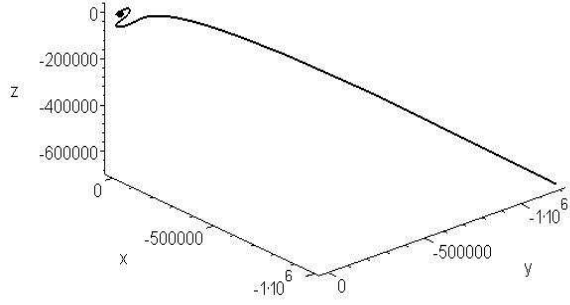}
	\end{center}
	\caption{The case $i_A=3$, $i_B=4$.}\label{sm6_34}	
\end{figure}
The integration is carried during the time range corresponding approximately to $t=3.23$ days.
The particle trajectory is shown in fig. \ref{sm6_34}.

In the case of unbounded motion, to define the integration interval firstly one has to find the nearest pole of $Q_1(\tau)$ and/or $Q_3(\tau)$ in the direction of ascending $\tau$.
Suppose this nearest pole is at $\tau=\tau_1$. Then we choose a small positive value $\varepsilon$ and divide the segment $[0, \tau_1-\varepsilon]$ into $N$ equal subsegments. The value $N $ is to be selected from practical reasons. 
The orbit should be visually smooth curve. In our examples the value $N=100 $ was used. After that, the calculations by the formulae derived above are carried out in equidistant nodes.

The following example demonstrates an application of our formulae for testing a numerical integration method. The original system of motion equations (\ref {K3}) is considered.
The Runge-Kutta-Fehlberg method of the eighth order with automatic choice of integration step is tested. 
The step is chosen by a method of the seventh order. The corresponding pair of programs, implemented in FORTRAN, is below noted as $RKF8(7) $.
 Integration of equations (\ref{K3}) by $RKF8(7) $ was performed with  relative local error of the method $ \varepsilon=10^{-13}$, and all calculations were carried out with double precision (real*8). The gravity parameter and the units of measurement are the same as above.  
A hypothetical particle is considered, repeatedly encountering the attracting centre. 
The trajectory obtained by explicit formulae is taken to be standard (reference). Its coordinates have been obtained using Maple with 32 digits (in FORTRAN this corresponds to quadruple precision (real*16)).
 
\prim Initial values of coordinates and velocities of a particle are as follows:
$$x_1=7000~\textrm{km},\quad x_2=0~\textrm{km}, \quad x_3=6000~\textrm{km},$$
$$\dot{x}_2 = 7.9~\textrm{km/s},\quad \dot{x}_1= \dot{x}_3=0~\textrm{km/s}
\quad (v_{cir}\approx 6.58~\textrm{km/s},\quad v_{par}\approx 9.30~\textrm{km/s}).$$
In an unperturbed case we have an elliptic  motion.

Parameters of retaining potential are as follows:
$$A_{-1}=0.1~\textrm{km}^4/s^2,\quad A_1=-0.02~\textrm{km}^2/s^2,\quad A_2=-0.2\cdot 10^{-5}~\textrm{km}/s^2$$
$$B_{-1}=-0.004~\textrm{km}^4/s^2,\quad B_1=-0.001~\textrm{km}^2/s^2,\quad B_2=-0.001~\textrm{km}/s^2$$
Coordinates of direction vector are $\widehat{{\bf b}}=(-1, -3, 1)^{T}$.
The roots of polynomials $\Phi_1$ and $\Phi_2$ are as follows:
$$\xi_2\approx 764,\quad \xi_3\approx 58639,\quad Q_1^0\approx  4459\quad
\Rightarrow Q_1^0 \in (\xi_2, \xi_3),$$ 
$$\eta_2\approx 504, \quad\eta_3\approx 7209,\quad Q_3^0\approx 4761 \quad
\Rightarrow Q_3^0 \in (\eta_2, \eta_3).$$
Therefore, the motion is bounded. The case  $i_A=3$, $i_B=3$.

\begin{figure}[h,t,p]
	\begin{center}
		\includegraphics[width=8cm ]{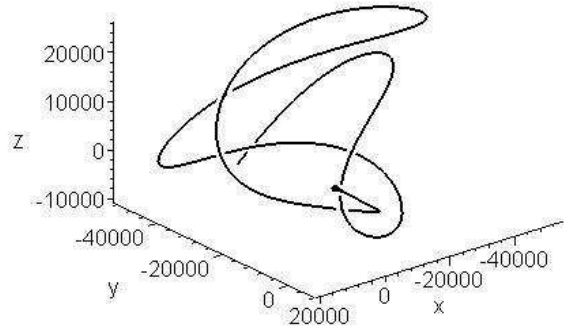}
	\end{center}
	\caption{The case $i_A=3$, $i_B=3$. The motion is bounded.}
	\label{so2_33_1_1000}	
\end{figure}

The calculations were carried out during the time ranges corresponding to 1, 10, 50, 100, 500, and 1000 revolutions of the particle around attracting centre without perturbations.  
The trajectory of the particle for three revolutions is shown in fig. \ref{so2_33_1_1000}. 

The table \ref{tab:Delta} contains the values, near the end of the integration interval, of the relative error for the energy constant $\delta H$, the coordinates of particle position vector $x_i$, and its absolute value $r$ 
$$\delta H =\frac{|H^0 - H |}{|H|},\quad \delta x_i =\frac{|x_i - x_i^c |}{|x_i|},\quad i=1,2,3, \quad \delta r =\frac{|r - r^c|}{r},$$ 
where $H^0$ is the value at the initial moment, $H$ at an arbitrary moment; 
$x_i^c$, $r^c$ are the values found by exact formulae. 
In the second column the intervals of physical time $t$ (in days) are given, for which numerical integration of system (\ref{K3}) was carried out.

 \begin{table}

	\caption{Estimation of the precision of numerical integration }
	\begin{center}
		\begin{tabular}{rrrrrrr}                        \hline
$n$   & $t$(day) &$\delta H\times10^{-12}$ &$\delta x_1\times10^{-12}$&$\delta x_2\times10^{-12}$ &$\delta x_3\times10^{-12}$ &$\delta r \times10^{-12}$      \\ \hline
1   &    .3382444  & 1    & 0.2     & 1       & 1      &  0.4     \\ 
10  &   4.9080991  & 2    & 6       & 12      & 213    &  10      \\ 
50  &  24.1940313  & 41   & 729     & 104     & 1667   &  108     \\ 
100 &  48.4322508  & 53   & 4399798 & 523748  & 95154  &  330606  \\ 
500 & 242.7821163  &294   & 77898   & 31418   & 151259 &  77206   \\ 
1000& 485.2955201  &556   & 554500  & 332688  & 1067003&  330900  \\ \hline			
		\end{tabular}      
	\end{center}
	\label{tab:Delta}
\end{table} 
From these data we can see that if the integration interval increases, the relative errors of $H$ and $x_3$ do not decrease. 
For coordinates $x_1$, $x_2$, and absolute value $r$, with $n=100$, these errors increase, then they diminish, and then increase again. 
 
The numerical examples show efficiency of the formulae we obtained. Besides, the theorem 3 allows to determine, given the initial position and velocity of a particle, whether its motion is bounded or unbounded.

\begin{center}
 \textbf{6. Conclusion}
\end{center}

In this paper we consider three sorts of coordinates (regular $q$-coordinates, bipolar coordinates, spherical coordinates). For each of the systems, the forms of potentials admitting complete separation of variables are given. Thus, the original equations for such potentials allow integration ``in the sense of Sundman''. 
In a similar way one can build, for regular $q$-coordinates, other coordinate systems for which Hamiltonian has orthogonal form, and with the use of Stackel theorem build potentials allowing the above-mentioned integrability.

Application of these potentials is a separate and independent problem. Those potentials are of practical worth which approximate some real forces.

\begin{center}
 \textbf{7. Acknowledgements}
\end{center}
The author is grateful to professor A.Zhubr for useful comments and discussions.


\begin{thebibliography}{99}
\bibitem{aksenov}
Aksenov, E.P.: 1977, \textit{Theory  of the motion of the Earth's artificial satellites.} 
Nauka, Moscow, 360 pp. (in Russian).
\bibitem{Fer}

 Ferrandiz, J.M. and Floria, L.: 1991, 'Towards a systematic definition of intermediaries in the theory of artificial satellites', \textit{Bull. Astron. Inst. Czechosl.},  \textbf{42},
  401 -- 407.                      
\bibitem{Belec}
	Beletskii, V.V.: 1964, 'Trajectories of Space Flights with a Constant Vector of Reactive Acceleration', \textit{Kosmicheskie Issledovaniya}, \textbf{2}(3), 787 -- 807. (in Russian).
 
\bibitem{Kunic}
Kunitsyn, A.L.: 1966, 'Rocket Motion in a Central Field of Forces with a Constant Vector of Reactive Acceleration', \textit{Kosmicheskie Issledovaniya},  \textbf{4}(2), 324 -- 332. (in Russian).
 
\bibitem{demin}
Demin, V.G.: 1968, \textit{The motion of  artificial satellite in the eccentric gravitational field}, Nauka, Moscow, 352 pp. (in Russian).

 
\bibitem{Kirch}
Kirchgraber, U.: 1971, 'A problem of orbital dynamics, which is separable in
$KS$-variables', \textit{Celest. Mech.} \textbf{4}, 340 -- 347.

\bibitem{PSM28}
Poleshchikov, S.M.: 2004, 'One integrable case of the perturbed two-body problem',
\textit{Cosmic Res}. \textbf{42}(4), 398 -- 407.

\bibitem{PSM21}
Poleshchikov, S.M. and Kholopov, A.A.: 1999,
\textit{Theory  of $L$-matrices and regularization of motion equations in 
Celestial Mechanics}, SFI, Syktyvkar,   255 pp. (in Russian).


\bibitem{PSM22}
Poleshchikov, S.M.: 2003, 'Regularization of motion equations with $L$-transformation and numerical integration of the regular equations',
 \textit{Celest. Mech. and Dyn. Astr.}   \textbf{85}(4), 341 -- 393.

\bibitem{Pars}
Pars, L.A.: 1965, \textit{A treatise on analytical dynamics}, Wiley,  NY, 636 pp.

\bibitem{Kholsh}
Kholshevnikov, K.V.: 1990, 'On the integrability in  celestial mechanics', 
\textit{Sbornik: Analitycal Celestial Mechanics}. Kazan. University,  5 -- 10. (in Russian)

\bibitem{KS}
Stiefel, E. and Scheifele, G.: 1971, \textit{Linear and regular celestial mechanics}, 
Springer-Verlag, Berlin,  304 pp.

\bibitem{PSM13}
Poleshchikov, S.M.: 1999, 'Regularization of canonical equations of the two-body
problem using a generalized $KS$-matrix', \textit{ Cosmic Res.} \textbf{37}(3), 302 -- 308.


\bibitem{PSM29}
Poleshchikov, S.M.: 2007, 'Motion of a particle in a perturbed field of the attracting centre', 
\textit{Cosmic Res.} \textbf{45}(6),  522 -- 535.


\bibitem{Byrd}
 Byrd, P.F. and Friedman, M.D.: 1954, \textit{Handbook of elliptic integrals for engineers and physicists}, Springer-Verlag, Berlin,  355 pp.

\bibitem{PSM30}
Poleshchikov, S.M. and Zhubr, A.V.: 2008, 'A set of potentials allowing integration of the perturbed two-body problem in regular coordinates', \textit{Cosmic Res}. (in press)


\bibitem{PSM31}
Poleshchikov, S.M.: 2006, 'An integrable case of the perturbed two-body problem producing elementary functions', \textit{Trudy SFI}, \textbf{ 6}, 31 -- 57. (in Russian)
\end{thebibliography}
\end{document}